\documentclass[11pt]{article}
\usepackage[letterpaper,margin=1in]{geometry}
\usepackage[T1]{fontenc}
\usepackage{authblk}

\usepackage[numbers,sort&compress]{natbib}
\usepackage{amsmath,amssymb,amsfonts}
\usepackage[ruled]{algorithm}
\usepackage{algpseudocode}
\makeatletter
\renewcommand{\fnum@algorithm}{\fname@algorithm~\thealgorithm:}
\makeatother
\usepackage{pgfplots}
\pgfplotsset{compat=1.16}
\usetikzlibrary{calc}
\usepackage{graphicx}
\usepackage{textcomp}
\usepackage{url}
\usepackage[hidelinks]{hyperref}

\title{A Human-Centric Adaptive Multi-Fidelity Digital Twin Framework for Cyber-Physical Systems}

\author[1]{Clarisa Grijalva-Lugo\thanks{Corresponding author: \href{mailto:clarisagl@arizona.edu}{clarisagl@arizona.edu}}}
\author[1]{Yu-Zheng Lin}
\author[1]{Shalaka Satam}
\author[2]{Sicong Shao}
\author[3]{Jesus Pacheco}
\author[1]{Soheil Salehi}
\author[1,4]{Pratik Satam}
\affil[1]{School of Electrical, Computing, and Software Engineering, The University of Arizona, Tucson, AZ, USA}
\affil[2]{School of Electrical Engineering and Computer Science, University of North Dakota, Grand Forks, ND, USA}
\affil[3]{Department of Industrial Engineering, University of Sonora, Hermosillo, Sonora, Mexico}
\affil[4]{Department of Systems and Industrial Engineering, The University of Arizona, Tucson, AZ, USA}
\date{}

\begin{document}

\maketitle

\begin{abstract}
Digital twins of cyber-physical systems must operate with scarce data, on assets that change after deployment, and for humans whose needs differ, yet existing implementations fix fidelity at design time and do not match the representation to the human or the task. This paper proposes a human-centric adaptive multi-fidelity digital twin framework that models the evolving human, the physical twin, and the digital twin core, and adds a human inference and adaptation function that estimates the human state, predicts the response, and adapts the representation. The framework defines low-, medium-, and high-fidelity tiers distinguished by adaptivity and by coupling, quantified as the ratio of the digital twin latency to the duration of the process step it follows. Three proof-of-concept instantiations are evaluated: a virtual reality tutor, a computer numerical control milling model, and an industrial training station. The low tier estimated learner sentiment with an accuracy of 0.86. The forecasters of the milling model and the training station, both medium-tier, showed a normalized error decline of 0.39 to 0.89 and a human adaptation gain of 1.16 to 1.51, but their latency exceeded the physical process step by 25 to 87 times. Only the station's vision model met the physical step. The tiers are separated by which functions adapt and where the latency constraint lies, not by a monotone increase in gain or coupling.
\end{abstract}


\section{Introduction}
\label{sec:introduction}
Advances in computing, networking, sensing, and artificial intelligence (AI) are redefining how cyber-physical systems (CPS) are designed and observed. In the transition to Industry 4.0, physical processes are treated as interconnected, observable, data-driven systems rather than isolated operational units \cite{ghobakhloo2020industry}. In a traditional Computer Numerical Control (CNC) milling process, for example, machine parameters are fixed in advance via G-code \cite{latif2021review}, and monitoring is limited to operator supervision or a few isolated machine signals. However, in an Industry 4.0 setting, the same process is observed through continuous data streams shared across networked infrastructure, enabling machines to coordinate with supervisory systems and scheduling platforms \cite{lins2017novel, garcia2022non, martinov2023approach}. This connectivity extends the value of process data beyond single-machine monitoring to production scheduling, predictive maintenance, and resource allocation across interconnected assets \cite{stojanovic2016big, qi2021fast, rossit2019industry, guo2021synchroperation, zonta2020predictive, mahmoodi2024data}.

Nevertheless, obtaining observable data that accurately reflects the physical system remains a challenge for CPS. Data on rare events, such as faults or attacks, are scarce and difficult to reproduce \cite{tang2025research}. Proprietary assets are protected as intellectual property \cite{stjepandic2015intellectual}, limiting knowledge of their firmware and design. In addition, physical assets change after deployment, so previously collected data will eventually diverge from the current state of the system \cite{li2021early}. These challenges motivate the use of \textit{digital twin (DT)} technology.

A DT is the computational counterpart of a physical system or process, referred to as a \textit{physical twin (PT),} that is continuously informed by observations from the physical world and capable of maintaining a behaviorally meaningful relationship with the system over time. The DT's relationship to the PT allows the user to inject faults and attacks to predict their effects, enabling the generation of data on rare events that cannot be readily reproduced on the PT. Continuous observation also permits the DT to track post-deployment changes in the PT, and its behavioral, rather than structural, relationship with the system removes the need to access undisclosed design knowledge.

Despite these capabilities, building a DT remains challenging in CPS and industrial environments. The twin must be built from observations of the PT, and unlike purely digital domains where data and computation scale at negligible cost, data acquisition is inseparable from physical operations. Gathering richer observations may require installing new sensors, modifying machine interfaces, or interrupting production for recalibration \cite{shi2023lstm, duro2016multi}. As a result, these settings are often data-scarce by design rather than merely data-limited in the statistical sense \cite{wuest2016machine}. A second constraint is the non-stationarity of deployed assets. Cumulative wear, degradation, and environmental variation alter asset behavior, so two nominally identical assets can diverge significantly under different operating histories \cite{li2021early}. A DT calibrated only to a system's initial nominal condition therefore loses validity over time, and any computational framework intended to support industrial decision-making must address not only modeling accuracy at a single point in time but also sustained adaptation to a changing physical system.

A third constraint concerns the people who use the DT. A technician, an engineer, and a researcher interacting with the same system may require fundamentally different information, abstraction levels, fidelity, and modalities for fault isolation, engineering design, or scientific investigation. The same DT, therefore, should adapt to its end user.

Existing DT approaches do not adequately address this combination of constraints. Prior implementations are often high- or low-fidelity models built on data-intensive pipelines \cite{zhao2023surrogate, ganti2020data} that are difficult to sustain under data scarcity. Multi-fidelity approaches combine or allocate fidelity to balance accuracy and cost, but do not align the representation with the human using the DT or with the task. This leaves a gap between the growing demand for DTs that remain accurate as data availability, system conditions, and user needs evolve, and the lack of a unified architecture spanning multiple fidelity levels, with instantiations at each level.

This paper addresses this gap by proposing HAMF-DT, a human-centric, adaptive, multi-fidelity DT framework for dynamic, data-driven, and adaptive intelligence in CPS, in which fidelity is treated as an explicit design dimension spanning low-, medium-, and high-fidelity representations rather than a fixed implementation choice.

The contributions of this paper are as follows:

\begin{enumerate}
\item A human-centric adaptive multi-fidelity DT framework, HAMF-DT, that models the evolving human, the PT, the DT core, and the application layer, and adds a human inference and adaptation function that estimates the human state, predicts the response, and adapts the representation.
\item A multi-fidelity taxonomy that distinguishes three tiers based on adaptivity and coupling and assigns a tier to each timed component.
\item Three instantiations of the framework, a virtual reality tutor (PRISM), a CNC milling model (DDD-GenDT), and an industrial training station (AIT-400), were evaluated for adaptation to the PT, adaptation to the human, and coupling.
\item A cross-tier comparison showing how the tiers differ in which functions adapt and where the latency constraint lies.
\end{enumerate}

The remainder of this paper is organized as follows. Section~\ref{sec:related-work} reviews related work on DT architectures, multi-fidelity modeling, and human-centered computing. Section~\ref{sec:reference-model} presents the DT reference model. Section~\ref{sec:framework} presents the proposed framework, and Section~\ref{sec:taxonomy} presents its multi-fidelity taxonomy. Sections~\ref{sec:low-fidelity}--\ref{sec:high-fidelity} present the low-, medium-, and high-fidelity cases. Section~\ref{sec:cross-case} consolidates findings across the three cases, and Section~\ref{sec:conclusion} concludes the paper.

\section{Related Work}
\label{sec:related-work}
\subsection{Digital Twin Architectures and Reference Models}
The DT concept originated in aerospace engineering as simulation-based systems engineering, in which high-fidelity DTs integrate multiphysics and multiscale simulations with sensor updates and historical data to continuously predict system behavior \cite{shafto2010draft}. Building high-fidelity behavioral physics models remains challenging, and the resulting simulations may not fully capture the physical behavior they represent \cite{li2024machine}. DTs have since spread to climate science, where they support weather forecasting and disaster response \cite{dale2023environment}; to medicine, where patient-specific models enable individualized diagnostics and treatment planning \cite{coorey2022health}; and to manufacturing, where they support real-time monitoring, predictive maintenance, and process optimization \cite{cimino2019review}.

The maturation of AI and increases in computing power have made ML-based modeling a practical alternative to physics-based simulation for developing DTs \cite{kaur2019convergence, rathore2021role, zhang2023lifetime}. In \cite{zhang2025process}, the authors proposed a multitask deep learning framework for monitoring tower pumping units under variable operating conditions. It jointly performs fault classification and regression of operating indicators and outperforms single-task and statistical baselines on real-world industrial data. This work treats DT development as a design space spanning multiple modeling paradigms and fidelity levels, rather than committing to a single fixed representation.

\subsection{Multi-Fidelity Modeling in Digital Twins}
Fidelity governs the alignment between a DT and its PT: higher fidelity requires denser sensing, finer spatiotemporal resolution, and more expensive simulation, while lower fidelity trades accuracy for lower cost and faster execution \cite{chen2022multi}. This tradeoff is exacerbated by data availability: high-fidelity sensor data is scarce, whereas complementary simulation data is more abundant but less accurate \cite{desai2023enhanced, wang2022building}. Multi-fidelity modeling addresses this by combining information sources of differing accuracy and cost, using two approaches: fusing the sources into a single surrogate model or embedding the choice of fidelity level within an active search or optimization loop \cite{chen2022multi}.

In the first approach, low- and high-fidelity data are fused into a single surrogate that approximates high-fidelity behavior at reduced cost. In \cite{desai2023enhanced}, Deep H-PCFE cascades hybrid polynomial correlated function expansion (H-PCFE) models that combine expansions with Gaussian processes, fusing data across fidelity levels via nonlinear autoregressive schemes where conventional linear fusion fails. In \cite{wang2022building}, a moving-least-squares-based multi-fidelity surrogate (MFS-MLS) reconstructs a continuous spatial field by calibrating dense but less accurate simulation output against sparse, accurate sensor measurements, and then trains a single-fidelity radial basis function surrogate on the calibrated field for real-time visualization.

In the second approach, the choice of which fidelity level to query and how much to trust it becomes part of an online optimization loop. In \cite{chen2022multi}, a convolutional multi-fidelity data-aggregation model is integrated into a neural optimization machine (MFNOM), so that low-fidelity simulation or scan data guides a stochastic-gradient-descent-based search over high-fidelity objectives for structural design and additive-manufacturing defect mitigation. In \cite{nobar2026guided}, a guided multi-fidelity Bayesian optimization framework (GMFBO) tunes closed-loop controllers by learning a localized Gaussian-process residual correction for state-dependent discrepancies between DT simulations and real measurements, using a cost-aware acquisition function that re-weights simulation trust as new data arrives. In both approaches, fidelity is combined or allocated to balance prediction accuracy and cost, but not as a function of the human using the DT or of the task.

\subsection{Human-Centered Computing}
Human-Centered Computing (HCC) organizes computing research around understanding humans, their activities, and their social and cultural context, and around designing systems that respond to them \cite{jaimes2006human, jaimes2007guest}. For DTs, this implies that the appropriate representation depends on the human interacting with the PT, not only on the PT. Existing research addresses this requirement through three approaches: human-centered AI, human-AI interaction, and human-centric DTs. Human-Centered AI treats automation and human control as independent design dimensions, with reliable, safe, and trustworthy systems requiring both to be high \cite{shneiderman2020human}. Human-AI interaction specifies how AI-generated information should be presented to people, using design guidelines validated with practitioners \cite{amershi2019guidelines} and findings on how people generate, select, and evaluate explanations \cite{miller2019explanation}. Human-centric DTs integrate human motion and biological sensing, AI, simulation, and augmented and virtual reality to support ergonomics, safety, training, and task allocation \cite{asad2023human}. These approaches adapt interactions, explanations, and content to the human, but do not jointly model the evolving human, PT, task, and context to select the DT representation.

\section{Digital Twin Reference Model}
\label{sec:reference-model}
This work adopts the DT reference model from \cite{lin2025ddd}. The model formalizes digital twinning as the continuous coupling of a \textit{physical space} and a \textit{virtual space} through data networks, communication protocols, and application programming interfaces (APIs). The physical space comprises the physical system and sensor data acquisition. The virtual space encompasses behavior modeling, interoperability, process management, and visualization. The model specifies five system-level indicators: security, accessibility, synchronization, interoperability, and fidelity.

\begin{figure}[htbp]
    \centering
    \includegraphics[width=0.97\linewidth]{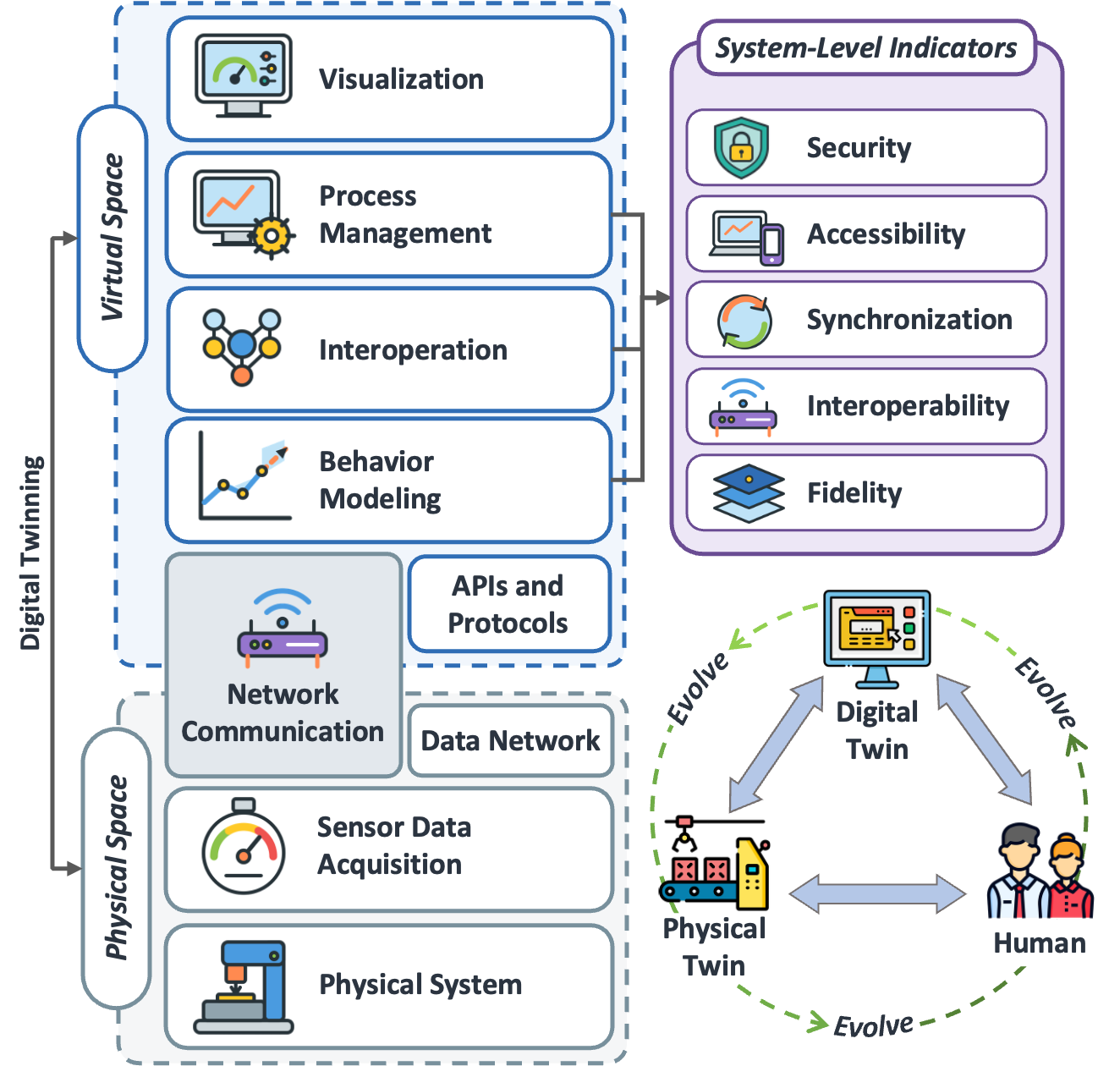}
    \caption{Digital twin reference model extended with the Human-Digital Twin-Physical Twin paradigm. Layers and system-level indicators adapted from \cite{lin2025ddd}.}
    \label{fig:digital-twin-reference-model}
\end{figure}

This work extends the reference model by introducing the human as a third entity, which establishes a Human-DT-PT paradigm, as shown in Fig. \ref{fig:digital-twin-reference-model}. The human interacts with both the DT and the PT, and the three entities evolve: the PT through operation, wear, and environmental change; the DT through adaptation to the PT and to the human, by adapting its information content and abstraction to the human's role and task; and the human through changes in goals, expertise, and information needs. Because the representation a human requires depends on the human's role, task, and operational context, which change over time, we treat fidelity in this work as a task-appropriate property rather than only as agreement between DT predictions and observed PT behavior, as defined in the reference model. Accordingly, this paper extends the reference model along the fidelity indicator, treating it as a design dimension organized into three tiers (Section~\ref{sec:taxonomy}).

\section{Human-Centric Adaptive Multi-Fidelity Digital Twin Framework}
\label{sec:framework}

\begin{figure}[htbp]
    \centering
    \includegraphics[width=\textwidth]{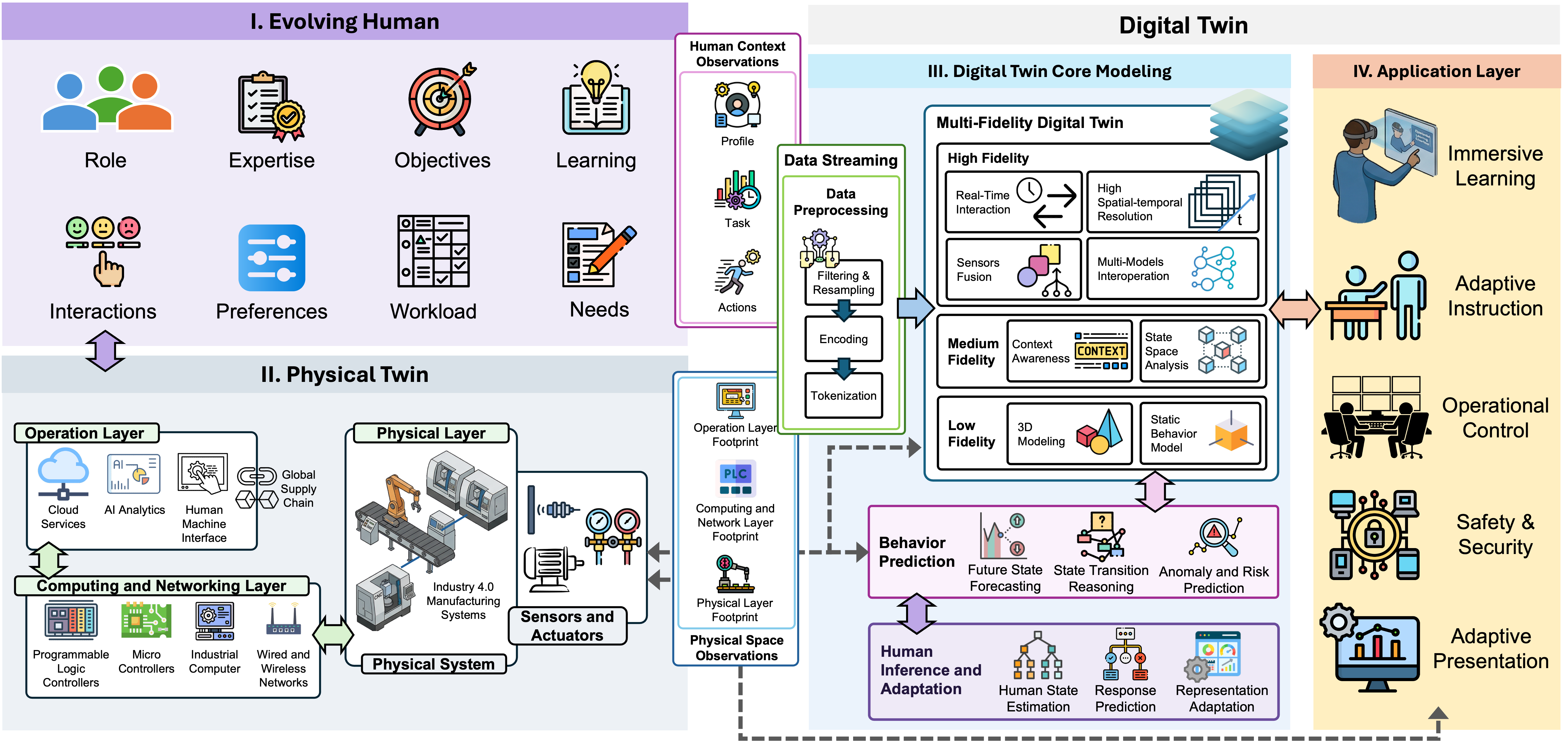}
    \caption{Human-centric adaptive multi-fidelity digital twin framework.}
    \label{fig:multi-fidelity-framework}
\end{figure}

This section presents the human-centric adaptive multi-fidelity DT framework (HAMF-DT), which organizes the DT as a bidirectionally coupled cyber-physical methodology spanning four components: the \textit{Evolving Human}, the \textit{Physical Twin}, the \textit{Digital Twin Core}, and an \textit{Application Layer}. The overall framework is shown in Fig. \ref{fig:multi-fidelity-framework}.

The framework emphasizes continuous interaction among the human, physical, and digital domains. The PT provides layered observations of the real system as physical, compute, network, and operational data, and the human provides observations of profile, task, and actions. The DT assimilates both into internal representations of varying fidelity. The resulting digital intelligence returns to the physical domain and to the human through feedback loops as improved situational awareness, intervention support, and adaptive operational response.

\subsection{Evolving Human}
The human is the entity for which the DT generates task-appropriate representations. Humans differ in the information, abstraction, and modality they require, and these requirements change as their goals, responsibilities, and expertise evolve. The framework therefore characterizes the human along eight categories:

\begin{itemize}
    \item \textit{Role:} Function and responsibilities of the human with respect to the system.
    \item \textit{Expertise:} Domain knowledge and skill of the human.
    \item \textit{Objectives:} The goals the human pursues with the system.
    \item \textit{Learning:} Acquisition of knowledge and skill that changes expertise over time.
    \item \textit{Interactions:} Manner in which the human engages with the system and its representations.
    \item \textit{Preferences:} The representation forms and interaction styles the human favors.
    \item \textit{Workload:} The demand placed on the human by the task.
    \item \textit{Needs:} The information and support the human requires.
\end{itemize}

These categories, however, are not observed directly. The framework accesses them through \textit{Human Context Observations}, grouped as profile, task, and actions, which enter the digital twin core through data streaming. Profile observations describe the human, task observations describe the activity being performed, and action observations record what the human does.

\subsection{Physical Twin}
The PT represents the real cyber-physical environment with its observable operational footprint. The proposed framework explicitly acknowledges that cyber-physical systems are typically monitored under heterogeneous sensing conditions, partial instrumentation, and evolving operational constraints. To address this, we introduce layered physical-space footprints, where each layer contributes complementary evidence about system behavior at a different level of abstraction. The footprints of the three layers together constitute the physical space observations.

\subsubsection{Operation layer}
The operation layer captures higher-level operational and service-oriented context that influences cyber-physical system behavior. Production goals, scheduling constraints, operator actions, service interactions, and external dependencies all affect how the physical and cyber layers should be interpreted. This layer links low-level observations to operational meaning, enabling the DT to move beyond isolated process monitoring toward contextually grounded, application-relevant interpretation.

\subsubsection{Computing and networking layer}
This layer captures the embedded cyber infrastructure that governs and coordinates physical operations, including industrial devices and wired/wireless communication networks. Because cyber-physical behavior depends on the computational logic and communication pathways that regulate the physical process, this layer complements the physical-layer footprint by exposing the underlying cyber mechanisms of system evolution.

\subsubsection{Physical layer}
The physical layer captures the underlying industrial process and its direct interaction with the physical world through sensors and actuators, reflecting process execution, equipment response, and measurable physical dynamics. It provides foundational evidence for describing the system's current condition. However, these observations are not always complete, noise-free, or temporally consistent. As a result, this layer is one essential footprint within a broader multi-layer observation space rather than a perfectly transparent representation of reality.

This layered view allows the DT to incorporate heterogeneous evidence without assuming that any single layer is sufficient on its own, treating the PT as a structured evidence space from which cross-layer relationships can be inferred.

\subsection{Digital Twin Core}
The DT is the methodological core of the proposed framework, transforming heterogeneous physical-space and human-context observations into structured representations. It is designed as a continuously updated inference space that assimilates incoming observations, interprets them under evolving operational and human context, and translates them into estimates of system behavior, future evolution, and human state. The DT bridges raw multi-layer evidence from the PT and the human and the downstream intelligence required for application-facing tasks.

\subsubsection{Data Streaming}
The DT core begins with a dynamic data-streaming pipeline that converts heterogeneous observations into machine-interpretable, temporally aligned representations. This transformation is performed through filtering, resampling, encoding, and tokenization. These operations establish temporal coherence and representation compatibility across modalities from different layers of the PT and from human-context observations.

\subsubsection{Multi-Fidelity Representation}
The DT core organizes its representations into three fidelity tiers, low, medium, and high, detailed in Section~\ref{sec:taxonomy}.

\subsubsection{Behavior Prediction}
The DT core translates inferred states and selected fidelity-level representations into actionable foresight through three predictive functions. \textit{1) Future-state forecasting} captures the system's short- or medium-horizon evolution under its current operating regime. \textit{2) State-transition reasoning} focuses on how the system moves between operational modes, process phases, or latent conditions. \textit{3) Anomaly and risk prediction} extends the DT from passive monitoring to proactive assessment by identifying deviations from expected behavior and estimating potential operational or security concerns before they fully manifest. 

\subsubsection{Human Inference and Adaptation}
The DT core infers the human's state from the Human Context Observations and adapts the representation it delivers to the human accordingly through three functions. \textit{1) Human state estimation} estimates the characteristics and current state of the human from profile, task, and action observations. \textit{2) Response prediction} predicts how the human will react to a candidate representation. \textit{3) Representation adaptation} adjusts the information content, abstraction, and level of detail of the representation as a function of the estimated human state and the predicted response. Algorithm~\ref{alg:human-adaptation} gives the sequence for interaction \(k\), with Human Context Observations \(o_k\), human state estimate \(\hat h_k\), candidate representations \(\mathcal{R}\), and predicted responses \(\hat y_k\).

\begin{algorithm}[t]
\caption{Human inference and adaptation at the DT core.}
\label{alg:human-adaptation}
\begin{algorithmic}[1]
\raggedright
\Require Human Context Observations \(o_k\) (profile, task, action); candidate representations \(\mathcal{R}\); initial human state estimate \(\hat h_0\)
\Ensure Delivered representation \(r_k^\ast\) for each interaction \(k\)
\For{\(k = 1, 2, \dots\)}
  \State \(\hat h_k \gets \textsc{Estimate}(o_k, \hat h_{k-1})\) \Comment{Human state estimation}
  \For{\(r \in \mathcal{R}\)}
    \State \(\hat y_k(r) \gets \textsc{Predict}(r, \hat h_k)\) \Comment{Response prediction}
  \EndFor
  \State \(r_k^\ast \gets \textsc{Adapt}(\mathcal{R}, \hat h_k, \hat y_k)\) \Comment{Representation adaptation}
  \State Deliver \(r_k^\ast\) to the human
\EndFor
\end{algorithmic}
\end{algorithm}

\subsection{Application Layer}
The application layer is the downstream interface through which the DT delivers actionable value to human users. This layer is designed as an extensible decision-support space that connects the DT core to five representative domains, and application outcomes and human responses return to the DT core through feedback loops.

\subsubsection{Immersive learning}
The DT functions as an experiential medium through which users observe, explore, and interact with cyber-physical processes. Because the DT maintains an internal representation grounded in physical-space observations and context-aware state reasoning, immersive environments can present what the system looks like, how it behaves, how it changes over time, and how different conditions affect operation. This makes the DT suitable for scenario-based learning, process walk-throughs, and exploratory training tasks.

\subsubsection{Adaptive instruction}
Here, learners interact with the DT within a guided instructional environment while large language models continuously interpret their natural-language responses to infer learner states (affective and pedagogical cues). These states are considered jointly with the DT's current state so instruction can be adapted in a context-sensitive manner.

\subsubsection{Operational control}
The application layer translates predictive insights into actionable decision support for supervision, intervention planning, and system coordination, assisting operators by exposing anticipated system trajectories, highlighting likely transitions between operational conditions, and providing early indications of emerging process deviations.

\subsubsection{Safety and security}
The DT also supports safety- and security-oriented monitoring by leveraging anomaly reasoning, risk interpretation, and contextualized knowledge support. This capability is complemented in two ways. First, it compares observed and predicted behavior to identify anomalous trajectories, suspicious control-flow deviations, or inconsistencies across physical, cyber, and operational layers. Second, it uses mined knowledge from large-scale technical corpora as an external reference layer that enriches interpretation of observed system behaviors.

\subsubsection{Adaptive presentation}
The DT adapts the form of its output, including the level of abstraction, modality, and detail, to the inferred state of the human and the task. Representation adaptation selects among output forms such as reports, dashboards, and immersive views, so that humans with different roles and expertise receive the same DT intelligence in the form their task requires.

\section{Multi-Fidelity Taxonomy}
\label{sec:taxonomy}
While the DT reference model aims to construct a high-fidelity virtual counterpart that accurately reflects the PT's dynamics, assuming universally maximal fidelity is neither practical nor necessary. Achieving extremely high fidelity requires dense sensing, fine-grained spatiotemporal resolution, and computationally intensive modeling, which incur high computational and energy costs and increase system complexity beyond operational requirements in many real-world deployments, hindering scalability and real-time responsiveness. Because the representation a human requires depends on the human's role, expertise, and objectives, a single maximal fidelity also cannot serve all humans. This section defines the three fidelity tiers, the two axes that distinguish them, and the rule for assigning a tier to a component.

\subsection{Fidelity Tiers}
To address this limitation, the framework organizes DT representations into a \textit{multi-fidelity DT reference model} spanning three fidelity levels: low, medium, and high. This allows model complexity to adapt to task requirements, available data, and computational constraints. Fidelity is thus treated as a dynamic attribute of the DT, jointly determined by the PT, the human, the task, and the operational context, enabling flexible transitions across representations of the same physical system. Table~\ref{tab:tiers} summarizes the tiers, which are described below.

\begin{table}[htbp]
    \centering
    \caption{Summary of the fidelity tiers by state space, adaptivity, coupling, and typical use}
    \footnotesize
    \renewcommand{\arraystretch}{1.5}
    \begin{tabular}{p{1.9cm}p{4.25cm}p{4.25cm}p{4.25cm}}
    \hline
     & \textbf{Low} & \textbf{Medium} & \textbf{High} \\
    \hline
    \textbf{State space} & Coarse-grained; variables for visualization and nominal behavior & Partially evolving; adds context from historical executions & High-dimensional; continuously updated by fused heterogeneous sensing \\
    \textbf{Adaptivity} & Trained offline; limited after deployment & Updates from new data without full retraining; triggered by data, events, or schedule & Continuously refined by incoming observations \\
    \textbf{Coupling} & No real-time coupling & Loose; analysis-driven; latency may exceed physical evolution & Tight; latency-constrained, bidirectional exchange \\
    \textbf{Typical use} & Visualization, baseline monitoring, offline analysis & Predictive analysis, cross-run knowledge transfer, adaptive monitoring, human adaptation & Real-time monitoring, control, safety-critical decisions \\
    \hline
    \end{tabular}
    \label{tab:tiers}
\end{table}

\subsubsection{Low-Fidelity}
At the lowest level, the DT provides an abstracted, computationally efficient representation of the PT, typically through 3D models or static behavior models. These models are trained offline and exhibit limited adaptability after deployment. The associated state space is coarse-grained, capturing only the variables required for basic visualization and approximation of nominal behavior. This level suits system visualization, baseline monitoring, and offline analysis, where computational efficiency is prioritized over continuous adaptability, and humans whose role and objectives are met by a coarse representation.

\subsubsection{Medium-Fidelity}
The medium-fidelity level introduces adaptive behavior modeling, where the DT moves beyond static representations and incorporates mechanisms to update its internal representation based on newly observed data. Medium-fidelity DTs maintain a partially evolving state representation whose state space expands to include contextual information from historical executions alongside instantaneous system variables. Through repeated assimilation of new data, the DT adapts to gradual changes such as wear, environmental variation, or operational drift without requiring full retraining. Adaptation is triggered by data availability, events, or scheduled intervals, rather than by real-time coupling, and inference latency may exceed the rate of physical system evolution. The DT therefore operates in a loosely coupled, analysis-driven mode, where updates inform future predictions rather than acting immediately on the physical system. This tier balances modeling flexibility against system-integration complexity and supports applications such as predictive analysis, cross-run knowledge transfer, adaptive system monitoring, and adaptation to the evolving human from repeated human responses.

\subsubsection{High-Fidelity}
At the highest level, the DT maintains a tightly coupled, dynamically synchronized representation of the PT, characterized by fine-grained spatial and temporal resolution \cite{fukami2023super} and continuous interaction with the PT. Here, the DT state space is high-dimensional and continuously updated by fusing heterogeneous sensing modalities into a coherent, temporally consistent view of the system. This enables capturing subtle variations and transient behaviors that lower-fidelity representations cannot capture. Additionally, the DT leverages interoperability across multiple modeling paradigms within a unified framework, allowing different models to operate collaboratively across scales and abstraction levels \cite{lahat2015multimodal, zhao2024deep}. The DT and PT interact through latency-constrained, bidirectional information exchange: incoming observations continuously refine the digital representation, and the updated DT provides timely feedback to the PT \cite{chen2025real, salzmann2023real}. This tightly integrated loop stays aligned with rapidly evolving system states and supports applications requiring high precision and responsiveness, such as real-time monitoring, control, and safety-critical decision-making, for the human tasks that require them.

\subsection{Tier-Diagnostic Axes}
\label{sec:tier-axes}
We distinguish the three tiers along two axes: \textit{adaptivity}, how much the internal representation updates in response to new data, and \textit{coupling}, the tightness and timing guarantees of the interaction loop between the DT and the PT. A tier is therefore not determined by the model class: a given class, such as a regression model or a physics-informed network, can serve in different tiers depending on how its representation adapts and how tightly it is coupled.

\subsubsection{Adaptivity}
We quantify adaptivity as the decline in forecast error as the DT assimilates new data. Let \(E_0\) be the forecast error at the first evaluated change and \(E_1\) the error at the end of the evaluation. The normalized error decline \(D\) is the reduction in error relative to \(E_0\).

\begin{equation}
    D = \frac{E_0 - E_1}{E_0}
    \label{eq:error-decline}
\end{equation}

A value of \(D > 0\) indicates that the forecast error declined over the evaluation, and \(D \le 0\) indicates that it did not. We report \(D\) together with the amount of task-specific training data used by the model. For the human loop, \(D\) is applied to the persona-estimate error and complemented by the adaptation gain \(G_H\), defined in Section~\ref{sec:medium-fidelity}.

\subsubsection{Coupling}
We quantify coupling as the ratio of the DT's latency to the duration of the physical process it follows. Let \(L_{\mathrm{DT}}\) be the wall-clock latency of one DT update or inference, measured from the start of input processing to the last output.

\begin{equation}
    L_{\mathrm{DT}} = t_{\mathrm{end}} - t_{\mathrm{start}}
    \label{eq:latency}
\end{equation}

Now, let \(\tau_{\mathrm{PT}}\) be the time the physical process takes to complete one step that the DT is expected to act on, and let \(\tau_H\) be the human response time scale, the time the human takes to respond to the DT’s output. Over \(N\) timed trials, the coupling ratio \(\rho\) is the median latency normalized by the time scale of the loop.

\begin{equation}
    \rho = \frac{\operatorname{median}(L_{\mathrm{DT},i})_{i=1}^N} {\tau_{\mathrm{PT}}}
    \label{eq:coupling-ratio}
\end{equation}

Finally, \(p_{\mathrm{exc}}\) is the fraction of trials in which the latency exceeds one process step. 

\begin{equation}
    p_{\mathrm{exc}} = \frac{1}{N} |\{i:L_{\mathrm{DT},i} > \tau_{\mathrm{PT}}\}|
    \label{eq:exceedance-rate}
\end{equation}

Replacing \(\tau_{\mathrm{PT}}\) with \(\tau_H\) gives \(\rho_H\) and \(p_{\mathrm{exc},H}\) for the human loop.

A value of \(\rho < 1\) indicates that a DT update completes within one process step, and \(\rho \ge 1\) indicates that updates lag the process.

\subsection{Tier Assignment}
The tier of a DT is assigned to each timed component and not to the system as a whole. The required tier of a component is determined by the human and the task it serves, and its supported tier is limited by its coupling. A system can therefore contain components of different tiers, and a tier is supported only by components whose coupling meets its requirements: a component with \(\rho \ge 1\) cannot support the high tier's latency-constrained, bidirectional exchange.

\section{Low-Fidelity: VR-Based Experiential Learning}
\label{sec:low-fidelity}
PRISM (Personalized, Rapid, and Immersive Skill Mastery) \cite{lin2025prism} is a virtual reality (VR) learning system for Industry 4.0 and cybersecurity workforce training. A learner works through low-fidelity VR modules featuring factory equipment and converses with a large language model (LLM) tutor via text or speech. A zero-shot LLM infers the learner's sentiment from the dialogue and task state, and that sentiment conditions a retrieval-augmented generation (RAG) pipeline \cite{raul2025rag}. The pipeline retrieves the top five 512-token chunks from a knowledge base index, and an LLM generates tutor guidance based on the learner's query and the retrieved chunks.

The VR lab is built offline, remains fixed after deployment, and is not fed by any live physical process. PRISM is therefore evaluated as a low-tier instance, not as a recommended tutoring method.

\subsection{Mapping to the HAMF-DT Framework}
PRISM provides the DT, the equipment in the VR lab stands in for the PT, and the learner is the Evolving Human. Table~\ref{tab:low-mapping} lists the elements of the HAMF-DT framework (Fig.~\ref{fig:multi-fidelity-framework}) that PRISM uses and how it implements each element.

\begin{table}[htbp]
    \centering
    \caption{Mapping of PRISM to the HAMF-DT framework}
    \footnotesize
    \renewcommand{\arraystretch}{1.5}
    \newcommand{\tbi}[1]{\hangindent=1em \hangafter=1 \noindent\textbullet~#1}
    \begin{tabular}{p{2.4cm}p{5.0cm}}
    \hline
    \textbf{HAMF-DT component} & \textbf{PRISM mapping} \\
    \hline
    \textit{Evolving Human} & Learner interacting with the tutor by text or speech \\
    \textit{Physical Twin} & Equipment represented in the VR lab, captured once by photogrammetry; no live physical data \\
    \textit{Human Context Observations} & \tbi{Actions: dialogue utterances}\par \tbi{Task: learning activities and completion patterns of the VR modules} \\
    \textit{Data Streaming} & Text sequences of dialogue and task state; query sequence of the learner query and sentiment vector \\
    \textit{Multi-Fidelity DT} & VR lab modules at the low tier, fixed after deployment; knowledge base index for retrieval \\
    \textit{Human Inference and Adaptation} & \tbi{Human state estimation: zero-shot LLM sentiment inference producing a sentiment vector}\par \tbi{Representation adaptation: sentiment-conditioned retrieval and LLM generation of tutor guidance} \\
    \textit{Application Layer} & \tbi{Immersive learning: four VR modules}\par \tbi{Adaptive instruction: tutor guidance delivered to the learner} \\
    \hline
    \end{tabular}
    \label{tab:low-mapping}
\end{table}

PRISM adapts to the human but not to the PT. The VR lab is fixed after deployment, so the DT does not assimilate PT data or forecast PT behavior. The DT observes the learner through dialogue and task state, and conditions tutor guidance on the sentiment inferred from the learner's utterances. Therefore, coupling is measured only for the human loop, as the DT's latency relative to the human step duration.

\subsection{Experimental Setup}
\subsubsection{Data and models}
The evaluations use the LLM-generated Education Dialogue Dataset \cite{shani2024multiturnreinforcementlearningpreference}. The EduTalk Sentiment Dataset consists of manually labeled conversations from this dataset, and its test set contains 1,289 conversations. Affect estimation uses one dialogue (topic: Personification, eight turn pairs), and the coupling measurement samples contexts from its eval split (7,234 conversations, 112,151 turns).

A seeded shuffle selected 25 topics and one conversation per topic. Each context is a prefix of 4 to 8 turns that ends on a student turn, so each model generates the next teacher turn. The same 25 contexts were used for every model. Retrieval used top-5 semantic search with \texttt{text-embedding-3-small} embeddings over a corpus of 38 chunks. Generation used a temperature of 0.7, an instruction to continue as the teacher in one short sentence, and no output-length cap. Five models were timed (GPT-3.5 Turbo, GPT-3.5 Turbo 16k, GPT-4, GPT-4 Turbo, and GPT-4.1). Each had 3 untimed warm-up calls followed by 25 sequentially issued timed calls, and no call failed.

\subsubsection{Human Inference and Adaptation}
The DT estimates the learner's sentiment via zero-shot LLM inference in two forms. In the classification, an LLM labels the conversation-level student sentiment of each transcript as positive or negative using a fixed prompt, without explanation. In the affect estimation \cite{lin2026llm}, each utterance is scored from 0 (most positive) to 5 (most negative) under a single rubric, and \(K = 20\) stochastic completions per utterance yield a distribution rather than a single label. The decoding temperature was swept from 0.1 to 1.0 with GPT-4.1, and six models (GPT-4.1, GPT-3.5 Turbo, Gemma 3 4B, Llama 3.3 70B, Phi-4 14B, and GPT-OSS 120B) were compared at temperature 0.7.

\subsubsection{Metrics}
The classification is scored using accuracy, precision, recall, specificity, and F1, with positive sentiment as the positive class. For affect estimation, scores are mapped to \([-1, 1]\) via \(\tilde{s} = 1 - 2(s/5)\), so that positive values indicate positive affect. The sample mean and variance, scaled by \((2/5)^2\), are computed per utterance. The learner trajectory is derived from the student turns (odd turn indices) in the per-utterance results reported in \cite{lin2026llm}, which analyzes the teacher trajectory in its stability figure.

Two human step durations are used because the DT acts after a learner turn and its output is consumed within a tutor turn: the learner step duration \(\tau_{H,\mathrm{learner}}\), the time to produce one learner turn, and the tutor step duration \(\tau_{H,\mathrm{tutor}}\), the time to deliver one tutor turn. In the eval split, student turns average 8.09 words and teacher turns 14.57 words. At a speech rate of 150 words per minute \cite{ncvs2024voice}, spoken turns take \(\tau_{H,\mathrm{learner}} = 3.2\) s and \(\tau_{H,\mathrm{tutor}} = 5.8\) s. At a typing rate of 52 words per minute \cite{dhakal2018typing}, a typed learner turn takes 9.3 s, and at a silent reading rate of 238 words per minute \cite{brysbaert2019many}, a read tutor turn takes 3.7 s. These values are derived from word counts rather than measured, because the dataset is LLM-generated and contains no timing metadata. The four values define the sensitivity range of the human time scale.

\(L_{\mathrm{DT}}\) is evaluated for the three components of the tutoring chain (sentiment inference, retrieval, and generation) and for the chain as their sum. Retrieval and generation were timed in a single streamed call, with the clock starting before retrieval and stopping at the final generated chunk. The time to first token (TTFT) was recorded from the same call.

Sentiment inference was not timed because four of the five models have been retired.\footnote{GPT-3.5 Turbo (\texttt{gpt-3.5-turbo-0125}), GPT-3.5 Turbo 16k, GPT-4, and GPT-4 Turbo were retired from the OpenAI API on 2026-10-23.} Latency per trial is estimated as TTFT minus retrieval latency, since the output is a single label and sentiment inference and generation are assumed to use the same model. The chain latency is the sum of this estimate and the measured retrieval and generation latencies. Affect estimation with \(K = 20\) completions per utterance is characterized by the latency headroom \(\tau - L_{\mathrm{DT}}\) it leaves.

\(\rho_H\) and \(p_{\mathrm{exc},H}\) are computed from \eqref{eq:coupling-ratio} and \eqref{eq:exceedance-rate} for each of the four step durations across the 25 trials. Intervals are the 95\% bootstrap interval of the median (10,000 resamples) divided by the step duration for \(\rho_H\), and the 95\% Wilson interval for \(p_{\mathrm{exc},H}\). A value of \(\rho_H\) is within one step when the median and the upper bound of its interval are below 1; it exceeds one step when the median and the lower bound are at least 1; otherwise, it is indeterminate.

The code for the experiments in this paper, including those in Sections~\ref{sec:medium-fidelity} and~\ref{sec:high-fidelity}, was written with the assistance of Claude Sonnet 5.0 and Claude Sonnet 5.5 (Anthropic) and was reviewed and tested by the authors.

\subsection{Experimental Analysis}
\subsubsection{Adaptation to the human}
GPT-4 achieved an accuracy of 0.86, a precision of 0.99, a recall of 0.84, a specificity of 0.97, and an F1 of 0.91 on the 1,289 test conversations. Of the positive conversations, 16\% were labeled negative, compared with 3\% of the negative conversations labeled positive, indicating that the classifier errs toward the negative class. The label is binary and carries no intensity, which the affect estimate provides.

Raising the decoding temperature from 0.1 to 1.0 increased the variance of the GPT-4.1 affect estimate for Utterance 6 from 0.010 to 0.024, while the sample mean remained between \(-0.11\) and \(-0.26\). For the learner, the mean at turn 3 stayed between \(-0.68\) and \(-0.60\), and the mean at turn 13 between 0.83 and 0.92, so the trajectory from a negative to a positive state was preserved. The mean affect estimate of GPT-4.1 for this dialogue is therefore stable under decoding stochasticity, while the variance retains the ambiguity of the utterance.

At temperature 0.7, GPT-4.1 and GPT-3.5 Turbo reproduced the early negative dip and the subsequent recovery of the trajectory, with a wider uncertainty band for GPT-3.5 Turbo. The four open-weight models showed a shallower or absent dip (nearly none for Llama 3.3 70B) or an incomplete recovery (approximately 0.40 for Phi-4 14B and 0.15 for Gemma 3 4B) \cite{lin2026llm}. Both analyses use a single synthetic dialogue.

The error decline \(D\) is not defined for PRISM because the twin does not assimilate data, and no per-step error series exists.

\subsubsection{Coupling}
The median chain latency ranged from 1891 ms for GPT-4.1 to 4336 ms for GPT-4 Turbo (Table~\ref{tab:low-coupling}). The interval of \(\rho_H\) was entirely below 1 in 17 of the 20 model-step-duration combinations. GPT-4 Turbo exceeded one learner step at the spoken duration (\(\rho_H = 1.34\)), whereas GPT-4 at that duration (0.95) and GPT-4 Turbo at the read tutor duration (1.18) were indeterminate.

\begin{table}[htbp]
    \centering
    \caption{Coupling of the PRISM tutoring chain to the human step durations}
    \footnotesize
    \renewcommand{\arraystretch}{1.2}
    \resizebox{\textwidth}{!}{%
    \begin{tabular}{lccccccc}
    \hline
    & & \multicolumn{4}{c}{\(\boldsymbol{\rho_H}\) \textbf{[95\% interval]}} & \multicolumn{2}{c}{\(\boldsymbol{p_{\mathrm{exc},H}}\)} \\
    \cline{3-6} \cline{7-8}
    & & \multicolumn{2}{c}{\textbf{Learner}} & \multicolumn{2}{c}{\textbf{Tutor}} & \textbf{Learner} & \textbf{Tutor} \\
    \cline{3-4} \cline{5-6}
    \textbf{Model} & \textbf{Median chain latency (ms)} & \textbf{Spoken (3.2 s)} & \textbf{Typed (9.3 s)} & \textbf{Spoken (5.8 s)} & \textbf{Read (3.7 s)} & \textbf{Spoken} & \textbf{Read} \\
    \hline
    GPT-3.5 Turbo      & 2542 & 0.79 [0.66, 0.99] & 0.27 [0.23, 0.34] & 0.44 [0.37, 0.55] & 0.69 [0.58, 0.87] & 0.28 & 0.16 \\
    GPT-3.5 Turbo 16k  & 2004 & 0.62 [0.54, 0.87] & 0.21 [0.19, 0.30] & 0.34 [0.30, 0.48] & 0.55 [0.47, 0.77] & 0.24 & 0.12 \\
    GPT-4              & 3064 & 0.95 [0.81, 1.07] & 0.33 [0.28, 0.37] & 0.53 [0.45, 0.59] & 0.83 [0.72, 0.94] & 0.40 & 0.24 \\
    GPT-4 Turbo        & 4336 & 1.34 [1.02, 1.66] & 0.46 [0.35, 0.58] & 0.74 [0.57, 0.92] & 1.18 [0.90, 1.46] & 0.84 & 0.60 \\
    GPT-4.1            & 1891 & 0.58 [0.52, 0.65] & 0.20 [0.18, 0.22] & 0.32 [0.29, 0.36] & 0.51 [0.46, 0.57] & 0.04 & 0.00 \\
    \hline
    \end{tabular}}
    \label{tab:low-coupling}
\end{table}

The retrieval and generation latency alone was within one learner step at the spoken duration for all models (\(\rho_H = 0.37\) to 0.81), so the estimated sentiment step moves the chain of GPT-4 from 0.58 to 0.95. Generation contributed the largest share of the chain at that duration (median \(\rho_H\) of 0.51 for GPT-4 and 0.26 for GPT-4.1), followed by sentiment inference (0.33 and 0.22) and retrieval (0.08 and 0.09).

For affect estimation with \(K = 20\) completions, the headroom of GPT-4.1 for sentiment inference, given the median retrieval and generation latency, is 2054, 8153, 4647, and 2491 ms for the spoken learner, typed learner, spoken tutor, and read tutor steps. Issued sequentially, each of the 20 completions would need to finish within 103, 408, 232, and 125 ms, respectively, whereas a single-label call is estimated at a median of 707 ms. Sequential issuance therefore fits none of the four step durations, and concurrent issuance would have to complete all 20 calls within the headroom.

The chain of GPT-4 Turbo exceeds one learner step at the spoken duration, and that of GPT-4 is indeterminate, so neither supports the latency-constrained exchange of the high tier at the pace of spoken learner turns. For GPT-4.1 and the GPT-3.5 variants, the interval of \(\rho_H\) lies below 1 at the four step durations, although 24\% to 28\% of the GPT-3.5 chain trials exceeded the spoken learner step. Because \(\rho_H < 1\) is not sufficient for a higher tier, the static VR lab and the single-pass sentiment conditioning, not the chain latency, place PRISM in the low tier.

The sentiment latency is an estimate, and speech-to-text, text-to-speech, and VR rendering were not timed, so the spoken \(\rho_H\) is a lower bound.

\section{Medium-Fidelity: Adaptive CNC Behavior Modeling}
\label{sec:medium-fidelity}
DDD-GenDT (Dynamic Data-Driven Generative Digital Twin) \cite{lin2025ddd} forecasts the behavior of a physical process from sensor data without task-specific training data. A Physical Twin Observation Graph (PTOG) organizes the measurements by run, process state, and sensor, and an observation window collects the same process state from the four preceding runs. After preprocessing, an ensemble of \(n = 10\) large language model (LLM) samples generated by zero-shot prompting forecasts the next process state, and the median across samples gives the point estimate. The original implementation used GPT-3.5 and GPT-4; here, we evaluate open-weight models. Health assessment computes the root mean square error (RMSE) between forecast and observation and applies a three-level policy (Continue, Warning, Stop) with thresholds \(T_{\mathrm{low}}\) and \(T_{\mathrm{high}}\). The result returns to the PT at Epoch Points, the locations on the process timeline at which a forecast is triggered.

The process is CNC milling from the NASA milling dataset \cite{agogino2007milling}, in which a milling tool cuts a workpiece over repeated runs while sensors record the process and the flank wear of the tool is measured. This work uses 14 runs and one sensor feed, the spindle motor current, which reflects both machining quality and tool wear. Each run is divided into three process states: P1, the initial cutting period, in which the current rises nonlinearly; P2, the stable cutting period, in which the current is approximately linear; and P3, the exit period. The spindle current is low-pass filtered at 8 Hz and downsampled to 20 Hz before it enters the PTOG. As the tool wears across runs, the current of the same process state changes, which the DT must track.

The zero-shot forecaster is the generative core of DDD-GenDT and is evaluated here as an instance of the medium tier, not as a recommended forecasting method. It needs no task-specific training data, because the four preceding runs are supplied in context. This property matters when no training set exists for a new asset.

\subsection{Mapping to the HAMF-DT Framework}
DDD-GenDT provides the DT, and the recorded CNC milling process stands in for the PT. Table~\ref{tab:medium-mapping} lists the elements of the HAMF-DT framework (Fig.~\ref{fig:multi-fidelity-framework}) that DDD-GenDT uses, and how it implements each one.

\begin{table}[htbp]
    \centering
    \caption{Mapping of DDD-GenDT CNC milling process to the HAMF-DT framework}
    \footnotesize
    \renewcommand{\arraystretch}{1.5}
    \newcommand{\tbi}[1]{\hangindent=1em \hangafter=1 \noindent\textbullet~#1}
    \begin{tabular}{p{2.4cm}p{5.0cm}}
    \hline
    \textbf{HAMF-DT component} & \textbf{DDD-GenDT CNC milling process mapping} \\
    \hline
    \textit{Evolving Human} & Simulated persona with experience, risk preference, and compliance; roles: operator, maintenance planner, and process engineer \\
    \textit{Physical Twin} & Physical layer footprint: CNC mill spindle current \\
    \textit{Human Context Observations} & Actions: accept or override, and response latency; profile and task observations are not available to the DT \\
    \textit{Data Streaming} & 8 Hz low-pass filter, downsampling to 20 Hz, encoding, and tokenization \\
    \textit{Multi-Fidelity DT} & \tbi{Context awareness: PTOG and four-run observation window}\par \tbi{State space analysis: LLM ensemble and median} \\
    \textit{Behavior Prediction} & \tbi{Future state forecasting: median forecast of the next process state}\par \tbi{Anomaly and risk prediction: health assessment} \\
    \textit{Human Inference and Adaptation} & \tbi{Human state estimation: posterior over persona parameters}\par \tbi{Response prediction: acceptance probability}\par \tbi{Representation adaptation: thresholds and explanation level} \\
    \textit{Application Layer} & \tbi{Operational control: health assessment decision (Continue, Warning, Stop)}\par \tbi{Adaptive presentation: alert, forecast, and explanation delivery and response capture} \\
    \hline
    \end{tabular}
    \label{tab:medium-mapping}
\end{table}

DDD-GenDT is designed to adapt to the PT and to the human. It adapts to the PT by accumulating the history of earlier runs, so that its forecasts follow the PT as the tool wears. The adaptation to the human is performed by updating its estimate of the human’s persona after each response, which sets the alert thresholds and explanation level it delivers. Both adaptations are evaluated separately, and coupling is measured for each loop as the latency of the DT relative to the time scale of what it follows: the process step \(\tau_{\mathrm{PT}}\) for the PT loop and the human response time scale \(\tau_H\) for the human loop.

The three entities exchange the following information. The DT observes the PT through spindle current windows and infers the next process state and the health state. It returns a health assessment decision (Continue, Warning, or Stop) to the PT at Epoch Points without altering the recorded observations. The DT delivers alerts, forecasts, and explanations to the human and observes only the human’s response, that is, whether an alert is accepted or overridden and the response latency. It starts with no prior information about the persona, infers the persona from these responses, and uses that inference to set its alert thresholds and explanation level.

\subsection{Experimental Setup}
\subsubsection{Data and forecasting}
The data are the spindle motor current of case 1 of the NASA milling dataset \cite{agogino2007milling}. Runs 1 to 4 serve as initial history, and runs 5 to 14 are forecast. Each run is divided into process states P1, P2, and P3 (100, 400, and 100 points) and into non-overlapping windows of 25 points, yielding 24 windows per run and 240 windows per model.

For each window, the history is the same-state window at the same offset from the four preceding runs (\(4 \times 25\) values), serialized digit by digit. The history depth and window length follow \cite{lin2025ddd} and were not varied. The model generates ten samples sequentially, which are combined by the elementwise median to give the forecast, and the forecast is scored by the RMSE against the observed window.

Four open-weight models were run without fine-tuning: Qwen2.5-14B, Qwen3-14B-Base, and Llama-3.1-8B as base models with temperature 0.7 (the GPT-3.5 configuration of \cite{lin2025ddd}), and Llama-3.1-8B-Instruct in chat mode with temperature 1.0 (the GPT-4 configuration), all with top-p 1.0. The models were served with vLLM on NVIDIA A10G GPUs.

\subsubsection{Simulated humans}
Each simulated human has a \textit{persona}, a tuple of three parameters: experience \(e\), risk preference \(r\), and compliance \(c\). The parameter \(r\) scales the alert threshold the human considers appropriate, so \(r < 1\) indicates a preference for earlier alerts. The parameter \(c\) sets the baseline tendency to accept an alert. Four personas combine novice (\(e = 0.3\)) or expert (\(e = 0.8\)) experience with cautious (\(r = 0.8\)) or risky (\(r = 1.25\)) risk preference: novice-cautious, novice-risky, expert-cautious, and expert-risky. By default, compliance is \(c = 0.6\). These results characterize the adaptation mechanism under assumed human models and not real human behavior. 

A persona accepts an alert with probability
\begin{equation}
\begin{split}
P(\mathrm{accept}) = \sigma\big(&\beta_0 + \beta_c\,\mathrm{logit}(c)\\
&+ \beta_m\, m + \beta_x\, x\,(1-e)\big)
\end{split}
\label{eq:accept}
\end{equation}
where \(\sigma\) is the logistic function, \(m = (S - r S_{\mathrm{ref}})/S_{\mathrm{ref}}\) is the margin of the alert statistic \(S\) over the persona's threshold \(r S_{\mathrm{ref}}\), and \(x \in \{0,1,2\}\) is the explanation level (minimal, standard, detailed). On the real-forecast stream, \(S\) is the window RMSE and \(S_{\mathrm{ref}}\) is the reference \(T_{\mathrm{low}}\). The term \(x\,(1-e)\) makes explanations raise acceptance most for novices. Explanations also slow the response, with
\begin{equation}
L_h = L_0 + \lambda\, x\,(1 + (1-e))
\label{eq:human-latency}
\end{equation}
and the observed latency carries log-normal noise with a log-scale standard deviation of 0.2. The constants \(\beta_0 = 0\), \(\beta_c = 2\), \(\beta_m = 1.5\), \(\beta_x = 1\), \(L_0 = 5\) s, and \(\lambda = 3\) s are assumed values fixed before the runs.

Three roles interact with the DT, and each accepts or rejects what the DT issues. The operator responds to each Warning or Stop alert. The planner responds to an inspection recommendation. The engineer responds to a recalibration proposal for \(T_{\mathrm{low}}\) and \(T_{\mathrm{high}}\), issued every two runs. The planner and the engineer have only 5 interactions each, so only acceptance and posterior width are reported for them. The planner and the engineer are not drifted. The planner's recommendation follows the run-level mean RMSE and a cumulative-RMSE check against \(T_{\mathrm{health}}\), the median cumulative RMSE over runs 5 to 14.

The operator's persona drifts at the midpoint of the alert stream. Three scenarios each vary a single parameter: 1) Experience growth, in which experience increases from 0.3 to 0.8; 2) Compliance drop, in which compliance falls from 0.6 to 0.3 in a step at the first false alarm after onset; and 3) Risk shift, in which risk preference ramps from \(r\) to \(1.4\,r\) to represent production pressure. A final combined scenario applies all three, and a no-drift control leaves the persona unchanged.

\subsubsection{Human Inference and Adaptation}
The adaptive DT implements the framework's three functions as follows. Human state estimation maintains a posterior over 27 persona hypotheses, with \(e \in \{0.3, 0.55, 0.8\}\), \(r \in \{0.8, 1.0, 1.25\}\), and \(c \in \{0.3, 0.6, 0.85\}\). After each decision \(a_k\) and observed latency \(L_k\), the weight of hypothesis \(\theta_i\) is updated as
\begin{equation}
w_i \leftarrow \frac{w_i^{\gamma}\, P(a_k, L_k \mid \theta_i, x, S)}{\sum_j w_j^{\gamma}\, P(a_k, L_k \mid \theta_j, x, S)}
\label{eq:posterior}
\end{equation}
with forgetting factor \(\gamma = 0.98\). The likelihood is the product of the acceptance probability in \eqref{eq:accept} and a log-normal density for the observed latency, and the weights start uniform. Response prediction evaluates \eqref{eq:accept} under the posterior. Representation adaptation sets \(T_{\mathrm{low}}\) and \(T_{\mathrm{high}}\) to their reference values scaled by the posterior mean \(\hat r\), and selects the explanation level that maximizes the expected acceptance minus a latency penalty of 0.005 per second of \(L_h\). Algorithm~\ref{alg:persona} summarizes the procedure on the real-forecast stream.

\begin{algorithm}[t]
\caption{Persona-based adaptation of the DT on the real-forecast stream.}
\label{alg:persona}
\begin{algorithmic}[1]
\raggedright
\Require Persona hypotheses \(\theta_1,\dots,\theta_{27}\); forgetting factor \(\gamma\); reference thresholds \(T_{\mathrm{low}}^{\mathrm{ref}}\) and \(T_{\mathrm{high}}^{\mathrm{ref}}\); latency penalty of 0.005 per second
\Ensure Alert level and explanation level for each alert
\State \(w_i \gets 1/27\) for all \(i\) \Comment{Initial weights}
\For{each window with alert statistic \(S\)}
  \State \(\hat r \gets \sum_i w_i r_i\) \Comment{Posterior mean risk preference}
  \State \(T_{\mathrm{low}} \gets \hat r\, T_{\mathrm{low}}^{\mathrm{ref}}\); \(T_{\mathrm{high}} \gets \hat r\, T_{\mathrm{high}}^{\mathrm{ref}}\)
  \If{\(S > T_{\mathrm{low}}\)}
    \For{\(x \in \{0,1,2\}\)}
      \State \(U(x) \gets \sum_i w_i\,[P(\mathrm{accept}\mid\theta_i,x) - 0.005\,L_h(\theta_i,x)]\) \Comment{\eqref{eq:accept}, \eqref{eq:human-latency}}
    \EndFor
    \State \(x^\ast \gets \arg\max_x U(x)\)
    \State \(\mathit{alert} \gets \mathrm{Warning}\) if \(S \le T_{\mathrm{high}}\), otherwise \(\mathrm{Stop}\); present the alert with \(x^\ast\)
    \State \((a_k, L_k) \gets\) observed decision and latency
    \State \(w_i \gets w_i^{\gamma}\, P(a_k, L_k \mid \theta_i, x^\ast, S)\) \Comment{\eqref{eq:posterior}}
    \State \(w_i \gets w_i \big/ \sum_j w_j\) \Comment{Normalization}
  \EndIf
\EndFor
\end{algorithmic}
\end{algorithm}

The non-adaptive baseline uses \(\hat r = 1\) and \(x = 1\), with a persona estimate fixed at the center of the grid (\(e = 0.55\), \(r = 1.0\), \(c = 0.6\)). Four conditions cross the forecast source with the policy: real forecast or oracle, and non-adaptive or adaptive. In the real-forecast conditions, \(T_{\mathrm{low}}\) and \(T_{\mathrm{high}}\) are the 75th and 95th percentiles of the 240-window RMSE values for each model, so an alert signals a deviation of the observation from the forecast, not tool wear. In the oracle conditions, the alert derives from the running maximum of flank wear VB: Healthy below 0.30 mm, Warning from 0.30 to 0.45 mm, and Stop at 0.45 mm or more. In the adaptive oracle condition, the Warning band becomes \(0.30\,\hat r\) (clipped at 0.449 mm) and the Stop band is unchanged, so adaptation cannot cause an unplanned stop by design. Accepting a Warning or Stop alert triggers an inspection that results in a tool change. The oracle conditions separate the effect of forecast error from the human layer.

\subsubsection{Metrics}
Adaptation to the PT is measured by the mean window RMSE of a run, \(\mathrm{Err}_{avg}\), pooled over its 24 windows, and by the normalized error decline \(D\) of \eqref{eq:error-decline} between run 5 (\(E_0\)) and run 14 (\(E_1\)). The published results of GPT-3.5, GPT-4, and a 1D-CNN autoencoder from \cite{lin2025ddd} serve as references.

Adaptation to the human is measured by five quantities. The persona-estimate error is
\begin{equation}
\epsilon_k = \frac{1}{3}\sum_{q \in \{e,\, r,\, c\}} \frac{\left|\hat\theta_{q,k} - \theta_{q,k}\right|}{R_q}
\label{eq:eps}
\end{equation}
where \(\hat\theta_{q,k}\) is the posterior mean, \(\theta_{q,k}\) the true value at interaction \(k\), and \(R_q\) the range of parameter \(q\) on the grid. The normalized decline \(D\) of \eqref{eq:error-decline} is applied to \(\epsilon_k\), with \(E_0\) the error at the first interaction and \(E_1\) the mean over the last ten interactions. The rate \(\alpha_H\) is the negative least-squares slope of \(\ln \epsilon_k\) over the interaction index, computed over the full stream and over the pre-onset phase. The time-to-adapt is the number of interactions after onset until \(\epsilon_k \le \max(1.1\,\bar\epsilon_{\mathrm{pre}}, 0.05)\) holds for ten consecutive interactions, where \(\bar\epsilon_{\mathrm{pre}}\) is the mean over the 20 interactions before onset; runs that never satisfy it are right-censored. The adaptation gain \(G_H\) is the ratio of the non-adaptive to the adaptive value for error-type metrics and of the adaptive to the non-adaptive value for acceptance-type metrics, so \(G_H > 1\) favors the adaptive policy.

Coupling to the PT uses \(L_{\mathrm{DT}}\), the wall-clock time for one window prediction (ten samples, parsing, and median), and \(\tau_{\mathrm{PT}} = 1.25\) s, the duration of one window (25 points at 20 Hz). Coupling to the human gives \(\rho_H\) and \(p_{\mathrm{exc},H}\) by \eqref{eq:coupling-ratio} and \eqref{eq:exceedance-rate} with \(\tau_H\). \(L_{\mathrm{DT}}\) is the latency of one window for the operator, the summed latency over a run for the planner, and the summed latency over a two-run block for the engineer. The values of \(\tau_H\) are assumed: 60 s, 30 min, and 2 h for the operator, planner, and engineer. \(\rho_H\) is also computed at 0.25 and 4 times these values.

Outcome quality is reported for the oracle conditions as tool utilization, the flank wear at the tool change divided by 0.45 mm, and as the fraction of seeds with an unplanned stop.

\subsection{Experimental Analysis}
\subsubsection{Adaptation to the PT}
\(\mathrm{Err}_{avg}\) decreased across all four open-weight models from run 5 (0.96 to 1.03) to run 14 (0.12 to 0.36), yielding \(D\) values between 0.65 and 0.88 (Table~\ref{tab:medium-adaptivity}). The GPT-3.5, GPT-4, and 1D-CNN autoencoder rows present published results from \cite{lin2025ddd}. Because the model weights are fixed, the decline reflects the DT tracking the worn tool's changing spindle current over the history of earlier runs in its observation window, not parameter updates. The decline is consistent with the adaptation reported for DDD-GenDT \cite{lin2025ddd}. The mean over runs 5 to 14 ranged from 0.47 for Qwen3-14B-Base to 0.65 for Llama-3.1-8B-Instruct, compared with 0.48 for the published GPT-4 and 0.96 for GPT-3.5. The autoencoder trained on runs 1 to 4 remains lower (mean 0.16) but declines little (\(D = 0.25\)), so the zero-shot models trade absolute accuracy for the absence of training data.

\begin{table}[htbp]
    \centering
    \caption{Adaptivity of DDD-GenDT to the PT}
    \footnotesize
    \renewcommand{\arraystretch}{1.2}
    \setlength{\tabcolsep}{4.5pt}
    \begin{tabular}{lcccc}
    \hline
    & \(\boldsymbol{\mathrm{Err}_{avg}}\) & \(\boldsymbol{\mathrm{Err}_{avg}}\) & \textbf{Mean} & \\
    \textbf{Model} & \textbf{Run 5} & \textbf{Run 14} & \textbf{runs 5--14} & \(\boldsymbol{D}\) \\
    \hline
    GPT-3.5 Turbo   & 2.79 & 0.26 & 0.96 & 0.91 \\
    GPT-4           & 0.95 & 0.24 & 0.48 & 0.75 \\
    1D-CNN autoencoder & 0.16 & 0.12 & 0.16 & 0.25 \\
    \hline
    Qwen2.5-14B                     & 0.99 & 0.18 & 0.51 & 0.82 \\
    Qwen3-14B-Base                  & 0.96 & 0.20 & 0.47 & 0.79 \\
    Llama-3.1-8B                    & 1.03 & 0.36 & 0.59 & 0.65 \\
    Llama-3.1-8B-Instruct           & 1.01 & 0.12 & 0.65 & 0.88 \\
    \hline
    \end{tabular}
    \label{tab:medium-adaptivity}
\end{table}

\subsubsection{Adaptation to the human}
The adaptive DT had a lower mean persona-estimate error than the non-adaptive baseline across all five scenarios (Table~\ref{tab:medium-human}). Without drift, the error was 0.226 versus 0.333, yielding adaptation gains \(G_H\) on \(\epsilon_k\) of 1.30 to 1.51. The table values are means over four models, four personas, and 30 seeds for the operator under real-forecast alerts. The gain on acceptance was 1.04 to 1.09. The gain on \(\epsilon_k\) is partly determined by the design, because the baseline holds a fixed estimate at the center of the grid, so its error is the distance of each persona from that estimate. The acceptance comparison does not have this property. The gain of a single model and persona ranged from 0.90 to 1.86.

\begin{table}[htbp]
    \centering
    \caption{Adaptation of the DT to the operator persona under real-forecast alerts}
    \footnotesize
    \renewcommand{\arraystretch}{1.2}
    \setlength{\tabcolsep}{4.5pt}
    \resizebox{\textwidth}{!}{%
    \begin{tabular}{lccccccc}
    \hline
    & \multicolumn{2}{c}{\textbf{Acceptance}} & \multicolumn{2}{c}{\textbf{Error} \(\boldsymbol{\bar\epsilon_k}\)} & \(\boldsymbol{D}\) \textbf{on} \(\boldsymbol{\epsilon_k}\) & \(\boldsymbol{\alpha_H}\) \textbf{(adaptive)} & \textbf{Time-to-adapt (adaptive)} \\
    \cline{2-3} \cline{4-5}
    \textbf{Scenario} & \textbf{Base / Adapt.} & \(\boldsymbol{G_H}\) & \textbf{Base / Adapt.} & \(\boldsymbol{G_H}\) & \textbf{Base / Adapt.} & \textbf{Pre / Full} & \textbf{Mean (censored share)} \\
    \hline
    No drift        & 0.855 / 0.886 & 1.04 & 0.333 / 0.226 & 1.49 &  0.00 / 0.51 & 0.020 / 0.013 & -- \\
    Experience growth & 0.849 / 0.882 & 1.04 & 0.315 / 0.244 & 1.33 &  0.12 / 0.29 & 0.021 / 0.007 & 6.9 (0.20) \\
    Risk shift      & 0.835 / 0.872 & 1.05 & 0.352 / 0.245 & 1.45 & -0.22 / 0.31 & 0.019 / 0.013 & 9.1 (0.31) \\
    Compliance drop & 0.617 / 0.672 & 1.09 & 0.424 / 0.283 & 1.51 & -0.55 / 0.28 & 0.021 / 0.003 & 23.5 (0.69) \\
    Combined        & 0.580 / 0.622 & 1.07 & 0.425 / 0.328 & 1.30 & -0.64 / -0.22 & 0.021 / -0.008 & 25.3 (0.77) \\
    \hline
    \end{tabular}}
    \label{tab:medium-human}
\end{table}

Without drift, \(D\) on \(\epsilon_k\) was 0.51 for the adaptive DT (from 0.36 at the first interaction to 0.18 over the last ten) and 0.00 for the fixed baseline estimate. The DT's persona estimate therefore converges toward the true persona as responses accumulate, whereas the fixed estimate does not. Under a single drifting parameter, the adaptive \(D\) was 0.28 to 0.31, and it became negative under combined drift (\(-0.22\)), where the error at the end of the stream exceeded the error at its start. The baseline \(D\) was negative under the risk shift, the compliance drop, and combined drift, so the adaptive DT tracked the moving persona while the fixed estimate did not. Under the combined drift, however, tracking fell behind the simultaneous changes.

The baseline \(\alpha_H\) ranged from \(-0.011\) to 0.004. For the adaptive DT, \(\alpha_H\) before onset was 0.019 to 0.021 per interaction, indicating that the error decreased by about 2\% per interaction. Over the full stream, \(\alpha_H\) was 0.013 without drift and under the risk shift, 0.007 under experience growth, 0.003 under the compliance drop, and \(-0.008\) under combined drift. Drift offsets part of the error reduction, and under combined drift, the error grew overall.

Time-to-adapt was 6.9 interactions for experience growth and 9.1 for the risk shift. Under the compliance drop and combined drift, the estimate did not re-enter the tolerance band before the end of the stream in 69\% and 77\% of the runs, respectively. Censored runs are counted at the remaining stream length, so time-to-adapt is a lower bound and is not defined without drift.

The posterior of risk preference stayed at its prior, so threshold adaptation had minimal effect.

Sensitivity analyses did not change the conclusions. A longer memory (larger \(\gamma\), from 0.90 to 1.00) lowered \(\bar\epsilon_k\) without drift (0.278 to 0.213) and raised it under combined drift (0.287 to 0.346), with the censored share of time-to-adapt rising from 0.41 to 0.90. At an initial compliance of 0.3, the acceptance rate was 0.478 for the adaptive DT compared with 0.375 for the baseline, and at 0.85 it was 0.984 against 0.987.

The planner and the engineer each had only 5 interactions, which is too few to measure the error decline. The adaptive acceptance rate was 0.789 for the planner (baseline 0.724) and 0.565 for the engineer (baseline 0.452). After the last interaction, the posterior standard deviation was 0.13 for experience and 0.11 to 0.13 for compliance, below the prior values of about 0.20 and 0.22, and 0.18 for risk preference, which remained unchanged from the prior.

\subsubsection{Coupling}
The median window latency \(L_{\mathrm{DT}}\) was 30.9 s for Llama-3.1-8B-Instruct and 68.4 to 70.9 s for the other three models, yielding \(\rho\) values of 24.8 to 56.7 relative to \(\tau_{\mathrm{PT}}\) (Table~\ref{tab:medium-coupling}). All 240 windows across every model exceeded \(\tau_{\mathrm{PT}}\) (\(p_{\mathrm{exc}} = 1.00\)), so no prediction completed within the 1.25 s of the process it forecasts, and the DT cannot act on the process in real time. This is consistent with the loosely coupled, analysis-driven interaction of the medium tier, in which updates inform later forecasts.

The human loop is less constrained than the PT loop. At the assumed \(\tau_H\) of 60 s, 30 min, and 2 h for the operator, planner, and engineer, \(\rho_H\) was 1.14 to 1.18 for the operator with the three slower models and 0.52 with Llama-3.1-8B-Instruct, with \(p_{\mathrm{exc},H}\) of 1.00 and 0.00. For the planner and the engineer, \(\rho_H\) was below 1 for every model, with \(p_{\mathrm{exc},H}\) of 0.00. The operator is therefore the only role whose DT update lags the human, and only for the three slower models. At 0.25 times \(\tau_H\), \(\rho_H\) exceeded 1 for every model and role except the Llama-3.1-8B-Instruct engineer, and at 4 times it was 0.30 or below, so the conclusions depend on the assumed \(\tau_H\).

\begin{table}[htbp]
    \centering
    \caption{Coupling of DDD-GenDT to the PT and to the human}
    \footnotesize
    \renewcommand{\arraystretch}{1.2}
    \setlength{\tabcolsep}{3pt}
    \begin{tabular}{lcccccc}
    \hline
    & \textbf{Median} & & \multicolumn{3}{c}{\(\boldsymbol{\rho_H}\)} & \(\boldsymbol{p_{\mathrm{exc},H}}\) \\
    \cline{4-6}
    \textbf{Model} & \(\boldsymbol{L_{\mathrm{DT}}}\) \textbf{(s)} & \(\boldsymbol{\rho}\) & \textbf{Op.} & \textbf{Plan.} & \textbf{Eng.} & \textbf{Op.} \\
    \hline
    Qwen2.5-14B           & 70.0 & 56.0 & 1.17 & 0.94 & 0.47 & 1.00 \\
    Qwen3-14B-Base        & 68.4 & 54.7 & 1.14 & 0.91 & 0.46 & 1.00 \\
    Llama-3.1-8B          & 70.9 & 56.7 & 1.18 & 0.95 & 0.47 & 1.00 \\
    Llama-3.1-8B-Instruct & 30.9 & 24.8 & 0.52 & 0.40 & 0.20 & 0.00 \\
    \hline
    \end{tabular}
    \label{tab:medium-coupling}
\end{table}

\subsubsection{Outcome quality}
In the oracle conditions, no seed had an unplanned stop under either policy. Values are means over four models, four personas, and 30 seeds. The adaptive DT lowered the mean tool utilization from 0.845 to 0.787 across the scenarios, which means that tools were changed earlier, at lower flank wear. The reduction was small without drift and under experience growth and the risk shift (0.844 to between 0.813 and 0.817), and larger under the compliance drop (0.749) and combined drift (0.742). The non-adaptive operator accepted the first Warning alert at a flank wear of 0.38 mm in run 11, whereas the adaptive Warning band \(0.30\,\hat r\) moved the first alert and the tool change to 0.33 to 0.37 mm. Under the compliance drop and combined drift, the adaptive DT needed 3.3 and 3.9 interactions on average to reach a change, against 3.7 and 4.9 for the baseline. These differences follow from the shifted Warning band and the response probabilities and do not indicate a learned improvement, because no seed reached the wear limit under either policy. The outcome results rest on a single tool life.

\section{High-Fidelity: Industrial Training Station}
\label{sec:high-fidelity}
The AIT-400 Artificial Intelligence Industrial Trainer \cite{smc2026ait400} is an industrial station that teaches artificial intelligence through manufacturing automation. A tower of test products is moved by electric actuators along linear axes. A programmable vision camera and a programmable logic controller (PLC) with OPC UA provide sensing and control, and an integrated computer performs model training and inference.

This work uses two AI processes of the station. Recorded actuator force traces are forecast using DDD-GenDT, and a convolutional neural network (CNN) performs tower recognition and provides the vision component. The trainee and the instructor are simulated.

The station is evaluated as a candidate high-fidelity DT. Only the vision model meets the coupling requirements for the high tier, and the forecaster is classified as medium-tier.

\subsection{Mapping to the HAMF-DT Framework}
The DT consists of the forecaster and the vision model, and the PT is represented by recorded runs of the AIT-400. The trainee and the instructor constitute the Evolving Human. Table~\ref{tab:high-mapping} lists the elements of the HAMF-DT framework (Fig.~\ref{fig:multi-fidelity-framework}) used by the AIT-400 and how it implements each one.

\begin{table}[htbp]
    \centering
    \caption{Mapping of the AIT-400 to the HAMF-DT framework}
    \footnotesize
    \renewcommand{\arraystretch}{1.5}
    \newcommand{\tbi}[1]{\hangindent=1em \hangafter=1 \noindent\textbullet~#1}
    \begin{tabular}{p{2.4cm}p{5.0cm}}
    \hline
    \textbf{HAMF-DT component} & \textbf{AIT-400 mapping} \\
    \hline
    \textit{Evolving Human} & Simulated persona with experience, risk preference, and compliance; roles: trainee and instructor \\
    \textit{Physical Twin} & Physical layer footprint: actuator force and tower images \\
    \textit{Human Context Observations} & Actions: accept or override, and response latency; profile and task observations are not available to the DT \\
    \textit{Data Streaming} & Moving-average filter, downsampling by 10 to 30 ms spacing, encoding, and tokenization \\
    \textit{Multi-Fidelity DT} & \tbi{Context awareness: setpoint groups and four-run observation window}\par \tbi{State space analysis: LLM ensemble and median}\par \tbi{Vision model: CNN tower recognition} \\
    \textit{Behavior Prediction} & \tbi{Future state forecasting: median forecast of the next force window}\par \tbi{Anomaly and risk prediction: health assessment} \\
    \textit{Human Inference and Adaptation} & \tbi{Human state estimation: posterior over persona parameters}\par \tbi{Response prediction: acceptance probability}\par \tbi{Representation adaptation: thresholds and explanation level} \\
    \textit{Application Layer} & \tbi{Operational control: health assessment decision (Continue, Warning, Stop)}\par \tbi{Adaptive presentation: alert, forecast, and explanation delivery and response capture} \\
    \hline
    \end{tabular}
    \label{tab:high-mapping}
\end{table}

The DT is designed to adapt to the PT and to the human. It adapts to the PT by using the four preceding runs in its observation window, so that its forecasts follow a regime change without retraining. The adaptation to the human is performed by updating its estimate of the human’s persona after each response, which sets the alert thresholds and explanation level it delivers. Both adaptations are evaluated separately, and coupling is measured for each loop as the DT’s latency relative to the time scale of what it follows: the move duration \(\tau_{\mathrm{PT}}\) for the PT loop and the human response time scale \(\tau_H\) for the human loop.

The DT observes the PT through force windows and infers the next window and the health state. It returns a health assessment decision (Continue, Warning, or Stop) to the PT without altering the recorded observations. The DT delivers alerts, forecasts, and explanations to the human and observes only the human’s response, including whether an alert is accepted or overridden and the response latency.

\subsection{Experimental Setup}
\subsubsection{Data and forecasting}
The data consist of 65 recorded runs of a single actuator move, grouped into three sets that share setpoints (speed 1100; acceleration and deceleration 800, 1000, or 1500), with 24, 20, and 21 runs ordered by date. The signal is the measured force (545 samples at 3 ms per move), smoothed with a moving average and downsampled by 10, yielding 30 ms spacing. Each target run, from the fifth run of its group onward, is divided into non-overlapping windows of 25 points (0.75 s). Only the first two windows are valid for every target because some history runs have fewer than 75 points, yielding 53 targets and 106 windows per model.

In each group, the final contiguous block of runs is labeled anomalous by the testbed operators, starting at the 15th, 11th, and 13th runs, making 29 of the 53 targets anomalous.

For each window, the history consists of the same window from the four preceding runs of the same group (\(4 \times 25\) values), serialized digit by digit. The model generates ten samples, which are combined using the elementwise median to produce the forecast. The forecast is then scored using RMSE against the observed window.

The four open-weight models in Section~\ref{sec:medium-fidelity} were run without fine-tuning, using the same configurations and serving setup on NVIDIA A10G GPUs. A previous-run copy, the mean of the four history runs, a ridge regression, and an ARIMA model serve as references across the same 106 windows. The ridge regression was trained on 48 pre-change windows and was not retrained after the change.

\subsubsection{Simulated humans}
The four personas, the acceptance model \eqref{eq:accept}, the response latency model \eqref{eq:human-latency}, and their constants are those from Section~\ref{sec:medium-fidelity}, with compliance \(c = 0.6\) by default. These results characterize the adaptation mechanism under assumed human models, not real human behavior.

Two roles interact with the DT, and each accepts or rejects the DT's outputs. The trainee responds to each Warning or Stop alert, yielding 34 to 54 interactions per model. The instructor responds to an inspection recommendation at each of the 53 targets. The recommendation is based on the target-level mean RMSE and a cumulative-RMSE check against \(T_{\mathrm{health}}\), the 75th percentile of the pre-change cumulative RMSE. The instructor is not drifted, so only acceptance and posterior width are reported for the instructor.

The five scenarios in Section~\ref{sec:medium-fidelity} (experience growth, compliance drop, risk shift, their combination, and a no-drift control) are applied to the trainee using the same parameter changes. Drift is applied starting at the middle alert of the baseline’s alert stream, at window 55 for Qwen2.5-14B and Qwen3-14B-Base, 57 for Llama-3.1-8B, and 74 for Llama-3.1-8B-Instruct. A false alarm that triggers the compliance drop is a Warning or Stop alert on a target that is not labeled anomalous.

\subsubsection{Human Inference and Adaptation}
The three functions are those of Section~\ref{sec:medium-fidelity}. Human state estimation maintains the posterior in \eqref{eq:posterior} over 27 persona hypotheses with a forgetting factor \(\gamma = 0.98\); response prediction evaluates \eqref{eq:accept} under the posterior; and representation adaptation scales \(T_{\mathrm{low}}\) and \(T_{\mathrm{high}}\) by the posterior mean \(\hat r\) and selects the explanation level with a latency penalty of 0.005 per second. The non-adaptive baseline and the four conditions, which cross the forecast source (real forecast or oracle) with the policy (non-adaptive or adaptive), are also those of Section~\ref{sec:medium-fidelity}.

In the real-forecast conditions, \(T_{\mathrm{low}}\) and \(T_{\mathrm{high}}\) are the 75th and 95th percentiles of the window RMSE for the 24 pre-change targets (48 windows) of each model. Calibration across all windows is not used because 29 of the 53 targets are anomalous, and pooled calibration produced 27 alerts in every model, below the minimum of 30 set before the runs. The alert signals a deviation of the observation from the forecast, not a fault, and by construction 12 of the 48 pre-change windows trigger an alert in every model.

In the oracle conditions, the alert is Continue before the first labeled target of a group and Warning from that point onward, because the label is binary and defines no Stop level. The alert is repeated at both windows of each target and at later targets of the block until it is accepted, and accepting it triggers an inspection. The oracle has no thresholds, so only the explanation level adapts. The oracle conditions separate the effect of forecast error from the human layer and form a ceiling case.

\subsubsection{Metrics}
Adaptation to the PT is measured by the normalized error decline \(D\) in \eqref{eq:error-decline}, where \(E_0\) is the error of the target run at the onset of the anomalous block and \(E_1\) is the mean error of the target runs four or more runs after the onset, averaged across the three groups. The error of a target run is the mean RMSE across its two windows, and the amount of task-specific training data is reported alongside \(D\).

Adaptation to the human is measured by the five quantities in Section~\ref{sec:medium-fidelity}: the persona-estimate error \(\epsilon_k\) of \eqref{eq:eps}, its normalized decline \(D\), the rate \(\alpha_H\), the time-to-adapt, and the adaptation gain \(G_H\). The pre-onset mean \(\bar\epsilon_{\mathrm{pre}}\) is computed over up to 20 interactions before onset, since the pre-onset phase has fewer than 20 interactions in two models.

Coupling to the PT uses \(L_{\mathrm{DT}}\), the latency of one move forecast (two sequential window forecasts, each of ten samples, with parsing and a median), and \(\tau_{\mathrm{PT}} = 1.64\) s, the duration of one move (545 samples at 3 ms). The median run-to-run cadence, 95.6 s, serves as a second reference. Coupling to the human gives \(\rho_H\) and \(p_{\mathrm{exc},H}\) via \eqref{eq:coupling-ratio} and \eqref{eq:exceedance-rate} with \(\tau_H\). For the trainee, \(L_{\mathrm{DT}}\) is the latency of one window, and \(\tau_H\) is the measured median run-to-run cadence, 95.6 s. For the instructor, \(L_{\mathrm{DT}}\) is the latency of one move forecast, and \(\tau_H\) is assumed to be 30 min. \(\rho_H\) is also computed at 0.25 and 4 times these values.

The latency of the tower-recognition CNN is the single-image inference time on a CPU, measured over 30 timed calls after 3 warm-up calls. The \(\tau_H\) of the vision path is assumed to be 1 s for an interactive response. Because the deployed checkpoint can only be scored in-sample, a CNN was retrained on an 80/20 stratified split of the 405 labeled images (seven classes), and its accuracy was computed on the 85 held-out images.

Outcome quality is reported for the oracle conditions as the detection delay, the number of targets from the first labeled target of a block to the accepted alert, the number of interactions per block, and the fraction of blocks without an accepted alert.

\subsection{Experimental Analysis}
\subsubsection{Adaptation to the PT}
All four open-weight models declined from the onset error to the late error, with \(D\) rising from 0.39 to 0.89 (Table~\ref{tab:high-adaptivity}). The weights are not updated, so the decline stems from the observation window, whose four preceding runs move past the onset and carry the new regime into the forecast. Qwen3-14B-Base matched ARIMA and fell slightly below the previous-run copy and the history mean, so the in-context tracking of the language models does not exceed that of classical baselines. The two Llama models declined least, and in one setpoint group each, \(D\) was near or below zero (\(-0.05\) for Llama-3.1-8B in group 1000 and 0.07 for Llama-3.1-8B-Instruct in group 1500).

Ridge, the only reference fitted to the data, declined less than the previous-run copy and the history mean (\(D\) of 0.71 against 0.92 and 0.91), so fitting on pre-change windows provided no advantage after the change.

\begin{table}[htbp]
    \centering
    \caption{Adaptation of the AIT-400 DT to the PT}
    \footnotesize
    \renewcommand{\arraystretch}{1.2}
    \begin{tabular}{lccc}
    \hline
    \textbf{Source} & \textbf{Mean RMSE} & \(\boldsymbol{E_1}\) & \(\boldsymbol{D}\) \\
    \hline
    Qwen2.5-14B & 6.89 & 5.04 & 0.67 \\
    Qwen3-14B-Base & 4.07 & 1.50 & 0.89 \\
    Llama-3.1-8B & 11.04 & 8.99 & 0.45 \\
    Llama-3.1-8B-Instruct & 10.06 & 9.06 & 0.39 \\
    \hline
    Previous-run copy & 1.59 & 1.06 & 0.92 \\
    History mean & 2.86 & 1.19 & 0.91 \\
    Ridge & 3.54 & 3.94 & 0.71 \\
    ARIMA & 3.05 & 1.42 & 0.89 \\
    \hline
    \end{tabular}
    \label{tab:high-adaptivity}
\end{table}

\subsubsection{Adaptation to the human}
The adaptive DT had a lower mean persona-estimate error than the non-adaptive baseline across all five scenarios (Table~\ref{tab:high-human}). The gain \(G_H\) on \(\epsilon_k\) ranged from 1.16 to 1.34, and the gain on acceptance went from 1.02 to 1.03. The table values are means over 30 seeds and the nine model and persona cells in which the experience and compliance parameters were recovered, for the trainee under real-forecast alerts. The gain for a single cell on \(\epsilon_k\) ranged from 0.91 to 1.51. Five of the nine cells are expert personas, for which the experience growth scenario equals the control.

\begin{table}[htbp]
    \centering
    \caption{Adaptation of the DT to the trainee persona under real-forecast alerts}
    \footnotesize
    \renewcommand{\arraystretch}{1.2}
    \setlength{\tabcolsep}{4.5pt}
    \resizebox{\textwidth}{!}{%
    \begin{tabular}{lccccccc}
    \hline
    & \multicolumn{2}{c}{\textbf{Acceptance}} & \multicolumn{2}{c}{\textbf{Error} \(\boldsymbol{\bar\epsilon_k}\)} & \(\boldsymbol{D}\) \textbf{on} \(\boldsymbol{\epsilon_k}\) & \(\boldsymbol{\alpha_H}\) \textbf{(adaptive)} & \textbf{Time-to-adapt (adaptive)} \\
    \cline{2-3} \cline{4-5}
    \textbf{Scenario} & \textbf{Base / Adapt.} & \(\boldsymbol{G_H}\) & \textbf{Base / Adapt.} & \(\boldsymbol{G_H}\) & \textbf{Base / Adapt.} & \textbf{Pre / Full} & \textbf{Mean (censored share)} \\
    \hline
    No drift          & 0.899 / 0.918 & 1.02 & 0.344 / 0.262 & 1.31 &  0.00 / 0.37 & 0.025 / 0.014 & -- \\
    Experience growth & 0.898 / 0.914 & 1.02 & 0.327 / 0.287 & 1.16 &  0.11 / 0.17 & 0.021 / 0.008 & 3.6 (0.16) \\
    Risk shift        & 0.893 / 0.910 & 1.02 & 0.401 / 0.327 & 1.23 & -0.53 / -0.17 & 0.022 / -0.002 & 13.0 (0.64) \\
    Compliance drop   & 0.799 / 0.824 & 1.03 & 0.411 / 0.307 & 1.34 & -0.53 / 0.05 & 0.023 / 0.000 & 10.0 (0.45) \\
    Combined          & 0.766 / 0.785 & 1.03 & 0.452 / 0.382 & 1.19 & -0.95 / -0.58 & 0.020 / -0.015 & 15.0 (0.72) \\
    \hline
    \end{tabular}}
    \label{tab:high-human}
\end{table}

Without drift, \(D\) on \(\epsilon_k\) was 0.37 for the adaptive DT and 0.00 for the fixed baseline estimate, indicating that the persona estimate converges toward the true persona as responses accumulate. Under experience growth and the compliance drop, the adaptive \(D\) fell to 0.17 and 0.05. It was negative under the risk shift (\(-0.17\)) and combined drift (\(-0.58\)), where the error at the end of the stream exceeded the error at its start. The baseline \(D\) was negative under all three of these scenarios (\(-0.53\), \(-0.53\), and \(-0.95\)), so the adaptive DT tracked the moving persona better than the fixed estimate, though not enough to reduce the error under risk shift or combined drift.

For the adaptive DT, \(\alpha_H\) before onset was 0.020 to 0.025 per interaction, a decrease of about 2\% in the error per interaction. Over the full stream, it was 0.014 without drift and 0.008 under experience growth, and it was zero or negative under the risk shift, the compliance drop, and combined drift (\(-0.002\), 0.000, and \(-0.015\)). Time-to-adapt was 3.6 interactions for experience growth, and the estimate did not re-enter the tolerance band before the end of the stream in 64\%, 45\%, and 72\% of the runs under the risk shift, the compliance drop, and combined drift, respectively.

The posterior of risk preference was not recovered in any cell, so threshold adaptation had minimal effect. A longer memory (larger \(\gamma\), from 0.90 to 1.00) lowered \(\bar\epsilon_k\) across all four models. At an initial compliance of 0.3, the gain in \(\epsilon_k\) was 2.23 to 2.27, and the gain in acceptance was 1.07 to 1.20; at 0.85, the gains were 1.47 to 1.55 and 1.00.

The instructor had 53 interactions with no drift, so only acceptance and the posterior width are reported. After the last interaction, the posterior standard deviation ranged from 0.02 to 0.06 for experience, 0.16 to 0.18 for risk preference, and 0.01 to 0.07 for compliance. The adaptive acceptance rate was 0.62 to 0.93, compared with 0.59 to 0.88 for the baseline, and 20 to 30 of the 53 interactions were recommended for inspection. The instructor results show that the posterior narrows with the instructor's responses, not that the DT improves the instructor's decisions: no drift was applied to the instructor, and the acceptance rates of the adaptive DT and the baseline overlap in range.

\subsubsection{Coupling}
The median move latency \(L_{\mathrm{DT}}\) was 64.8 s for Llama-3.1-8B-Instruct and 137.1 to 142.1 s for the other three models, yielding \(\rho\) values of 39.5 to 86.7 relative to \(\tau_{\mathrm{PT}}\) (Table~\ref{tab:high-coupling}). The median latency of every language model exceeded the 1.64 s move it forecasts, so the forecaster is a medium-tier component, usable between runs and not for interaction with the actuator. The tower-recognition CNN required 57.2 ms (\(\rho = 0.035\)), with a maximum of 86.9 ms in 30 calls, and is the only component whose coupling meets the latency-constrained exchange of the high tier. It achieved a held-out accuracy of 0.941 on 85 images (macro F1 0.937).

The human loop is less constrained than the PT loop. The window latency of the trainee path (the table lists the two-window move latency) was 68.7 to 70.9 s for the three slower models and 33.1 s for Llama-3.1-8B-Instruct, yielding \(\rho_H\) of 0.72 to 0.74 and 0.35 against the measured cadence of 95.6 s, with \(p_{\mathrm{exc},H}\) of 0.00. For the instructor at the assumed \(\tau_H\) of 30 min, \(\rho_H\) was 0.04 to 0.08, and for the vision path at the assumed 1 s it was 0.057. The forecaster therefore meets the human loop at the measured cadence and does not meet the loop of the actuator.

\begin{table}[htbp]
    \centering
    \caption{Coupling of the AIT-400 forecaster and vision model to the PT and to the human}
    \footnotesize
    \renewcommand{\arraystretch}{1.2}
    \setlength{\tabcolsep}{3pt}
    \begin{tabular}{lcccc}
    \hline
    & & & \multicolumn{2}{c}{\(\boldsymbol{\rho_H}\)} \\
    \cline{4-5}
    \rule{0pt}{3.4ex}\textbf{Source} & \shortstack{\textbf{Median}\\\(\boldsymbol{L_{\mathrm{DT}}}\) \textbf{(s)}} & \shortstack{\(\boldsymbol{\rho}\)\\\textbf{(1.64 s)}} & \shortstack{\textbf{Trainee}\\\textbf{(95.6 s)}} & \shortstack{\textbf{Instructor}\\\textbf{(30 min)}} \\
    \hline
    Qwen2.5-14B           & 140.7 & 85.8 & 0.74 & 0.08 \\
    Qwen3-14B-Base        & 137.1 & 83.6 & 0.72 & 0.08 \\
    Llama-3.1-8B          & 142.1 & 86.7 & 0.74 & 0.08 \\
    Llama-3.1-8B-Instruct & 64.8  & 39.5 & 0.35 & 0.04 \\
    \hline
    CNN                   & 0.057 & 0.035 & -- & -- \\
    \hline
    \end{tabular}
    \label{tab:high-coupling}
\end{table}

\subsubsection{Outcome quality}
In the oracle condition, the trainee accepted the alert at the first labeled window in nearly every block: without drift, no blocks were missed, the mean detection delay was 0 to 0.02 targets, and 1.0 to 1.1 interactions per block were needed. On the real-forecast stream, 0.51 to 0.90 of the 29 anomalous targets received an accepted alert, depending on the model (mean 0.67), because 32 of the 58 anomalous windows of Qwen2.5-14B and 36 of Llama-3.1-8B-Instruct remained below the Warning threshold. The alerts on pre-change targets numbered 11.0 to 12.0 per model, of which 9.9 to 11.2 were accepted; this number is fixed by calibration at the 75th percentile of the pre-change windows. The outcome results rest on one anomalous block per setpoint group and one run per target.

\section{Cross-Case Synthesis}
\label{sec:cross-case}

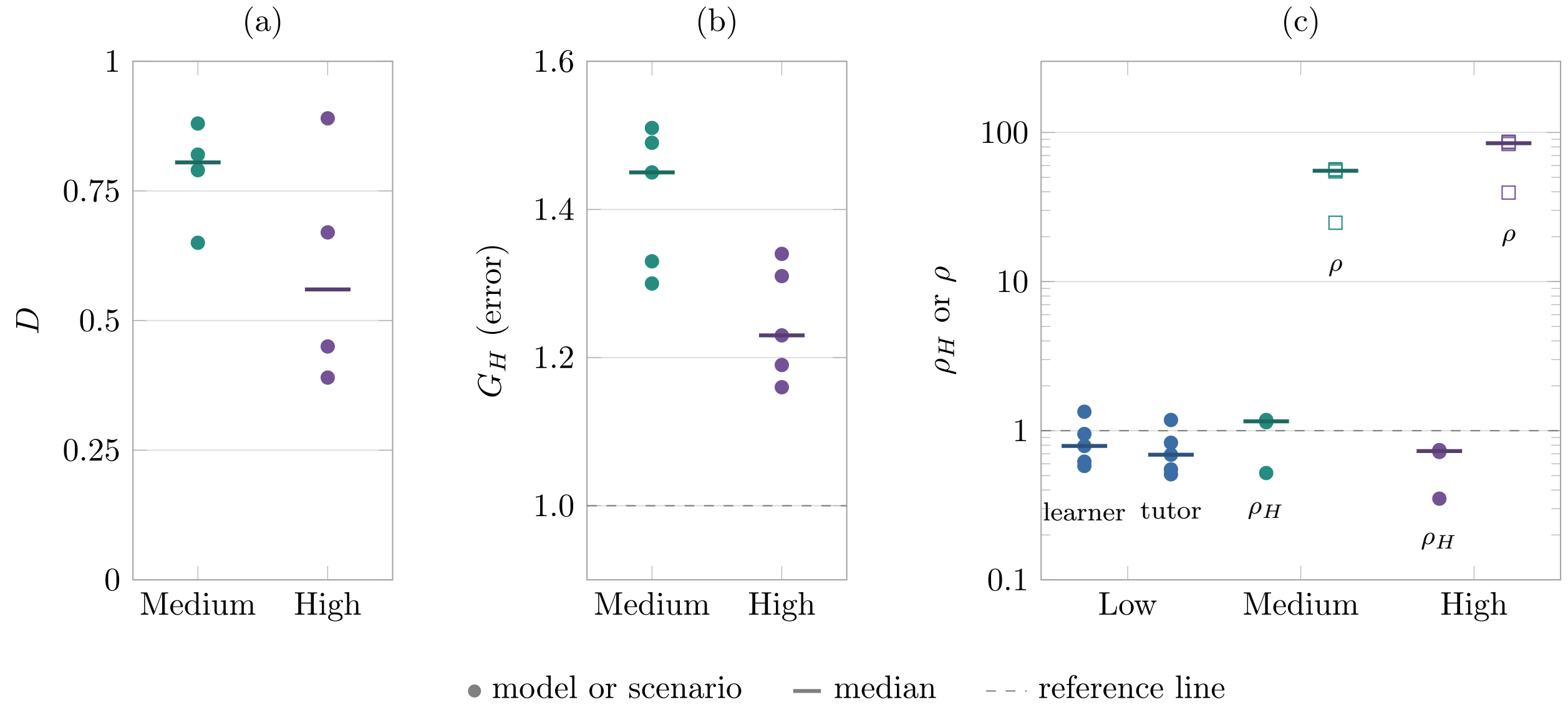
\begin{figure}[htbp]
\centering
\definecolor{tierL}{RGB}{58,110,165}
\definecolor{tierM}{RGB}{38,140,128}
\definecolor{tierH}{RGB}{115,82,150}
\pgfplotsset{tierplot/.style={scale only axis,height=5.6cm,xmin=0.5,xmax=2.5,xtick={1,2},xticklabels={Medium,High},tick label style={font=\small},label style={font=\small},title style={font=\small,yshift=-3pt},ymajorgrids,grid style={gray!25},axis line style={gray!70},tick style={gray!70}}}
\resizebox{\textwidth}{!}{%
\begin{tikzpicture}
\begin{axis}[tierplot,name=a,width=0.17\textwidth,title={(a)},ylabel={\(D\)},ymin=0,ymax=1,ytick={0,0.25,0.5,0.75,1},yticklabels={0,0.25,0.5,0.75,1}]
\addplot[only marks,mark=*,mark size=2pt,tierM] coordinates {(1,0.82)(1,0.79)(1,0.65)(1,0.88)};
\addplot[only marks,mark=-,mark size=7pt,very thick,tierM!75!black] coordinates {(1,0.805)};
\addplot[only marks,mark=*,mark size=2pt,tierH] coordinates {(2,0.67)(2,0.89)(2,0.45)(2,0.39)};
\addplot[only marks,mark=-,mark size=7pt,very thick,tierH!75!black] coordinates {(2,0.56)};
\end{axis}
\begin{axis}[tierplot,name=b,at={(a.east)},anchor=west,xshift=2.1cm,width=0.17\textwidth,title={(b)},ylabel={\(G_H\) (error)},ymin=0.9,ymax=1.6,ytick={1,1.2,1.4,1.6},yticklabels={1.0,1.2,1.4,1.6}]
\draw[gray,dashed] (axis cs:0.5,1) -- (axis cs:2.5,1);
\addplot[only marks,mark=*,mark size=2pt,tierM] coordinates {(1,1.49)(1,1.33)(1,1.45)(1,1.51)(1,1.30)};
\addplot[only marks,mark=-,mark size=7pt,very thick,tierM!75!black] coordinates {(1,1.45)};
\addplot[only marks,mark=*,mark size=2pt,tierH] coordinates {(2,1.31)(2,1.16)(2,1.23)(2,1.34)(2,1.19)};
\addplot[only marks,mark=-,mark size=7pt,very thick,tierH!75!black] coordinates {(2,1.23)};
\end{axis}
\begin{axis}[tierplot,name=c,at={(b.east)},anchor=west,xshift=2.1cm,width=0.34\textwidth,xmin=0.5,xmax=3.5,xtick={1,2,3},xticklabels={Low,Medium,High},title={(c)},ylabel={\(\rho_H\) or \(\rho\)},ymode=log,ymin=0.1,ymax=300,log ticks with fixed point,ytick={0.1,1,10,100}]
\draw[gray,dashed] (axis cs:0.5,1) -- (axis cs:3.5,1);
\addplot[only marks,mark=*,mark size=2pt,tierL] coordinates {(0.75,0.79)(0.75,0.62)(0.75,0.95)(0.75,1.34)(0.75,0.58)};
\addplot[only marks,mark=-,mark size=7pt,very thick,tierL!75!black] coordinates {(0.75,0.79)};
\addplot[only marks,mark=*,mark size=2pt,tierL] coordinates {(1.25,0.69)(1.25,0.55)(1.25,0.83)(1.25,1.18)(1.25,0.51)};
\addplot[only marks,mark=-,mark size=7pt,very thick,tierL!75!black] coordinates {(1.25,0.69)};
\node[font=\scriptsize,anchor=north] at (axis cs:0.75,0.4) {learner};
\node[font=\scriptsize,anchor=north] at (axis cs:1.25,0.4) {tutor};
\addplot[only marks,mark=*,mark size=2pt,tierM] coordinates {(1.8,1.17)(1.8,1.14)(1.8,1.18)(1.8,0.52)};
\addplot[only marks,mark=-,mark size=7pt,very thick,tierM!75!black] coordinates {(1.8,1.155)};
\addplot[only marks,mark=square,mark size=2pt,tierM] coordinates {(2.2,56.0)(2.2,54.7)(2.2,56.7)(2.2,24.8)};
\addplot[only marks,mark=-,mark size=7pt,very thick,tierM!75!black] coordinates {(2.2,55.35)};
\node[font=\scriptsize,anchor=north] at (axis cs:1.8,0.4) {\(\rho_H\)};
\node[font=\scriptsize,anchor=north] at (axis cs:2.2,17) {\(\rho\)};
\addplot[only marks,mark=*,mark size=2pt,tierH] coordinates {(2.8,0.74)(2.8,0.72)(2.8,0.74)(2.8,0.35)};
\addplot[only marks,mark=-,mark size=7pt,very thick,tierH!75!black] coordinates {(2.8,0.73)};
\addplot[only marks,mark=square,mark size=2pt,tierH] coordinates {(3.2,85.8)(3.2,83.6)(3.2,86.7)(3.2,39.5)};
\addplot[only marks,mark=-,mark size=7pt,very thick,tierH!75!black] coordinates {(3.2,84.7)};
\node[font=\scriptsize,anchor=north] at (axis cs:2.8,0.25) {\(\rho_H\)};
\node[font=\scriptsize,anchor=north] at (axis cs:3.2,27) {\(\rho\)};
\end{axis}
\node[anchor=north,font=\small,yshift=-0.9cm] at ($(a.south west)!0.5!(c.south east)$) {\tikz[baseline=-0.5ex]{\fill[gray] (0,0) circle (2pt);}~model or scenario\hspace{1.5em}\tikz[baseline=-0.5ex]{\draw[gray,very thick] (-0.15,0)--(0.15,0);}~median\hspace{1.5em}\tikz[baseline=-0.5ex]{\draw[gray,dashed] (-0.22,0)--(0.22,0);}~reference line};
\end{tikzpicture}}
\caption{Adaptation and coupling of the three tiers. (a) Normalized error decline \(D\) of the four open-weight models, not defined for PRISM. (b) Adaptation gain \(G_H\) on the persona error for each drift scenario, not applicable to PRISM. (c) Coupling ratio \(\rho_H\) of the human step (PRISM: learner and tutor; DDD-GenDT: operator; AIT-400: trainee) and \(\rho\) of the PT step, on a log axis. The plotting code was written with the assistance of Claude Sonnet 5.5 (Anthropic) from the measured values reported in Sections~\ref{sec:low-fidelity}--\ref{sec:high-fidelity}.}
\label{fig:cross-case}
\end{figure}

The three tiers differ in which DT functions adapt and in which latency constraint binds (Fig.~\ref{fig:cross-case}). The low-fidelity twin maintains a static PT representation and infers the learner state via zero-shot sentiment analysis, so neither \(D\) nor \(G_H\) is defined. The medium-fidelity twin and the forecaster of the high-fidelity station adapt to both the PT and the human, with larger gains in the medium tier (median \(D\) of 0.81 against 0.56, and median \(G_H\) of 1.45 against 1.23). The PT step binds for both forecasters, with DT latency 25 to 57 times the PT timescale in the medium tier and 40 to 87 times in the high tier, and only the vision model of the station meets it (\(\rho\) of 0.035). The human step is met in the high tier (\(\rho_H\) of 0.35 to 0.74) and exceeded by three of four models in the medium tier (1.14 to 1.18) and by one model in the low tier (GPT-4 Turbo). The tiers therefore differ by which functions adapt and where the latency constraint lies, not by a monotone increase in gain or coupling.

\subsection{Adaptation}
Adaptation to the PT was measured in the medium- and high-fidelity cases. The normalized error decline \(D\) for the four open-weight models ranged from 0.65 to 0.88 for DDD-GenDT (median 0.81) and from 0.39 to 0.89 for the AIT-400 (median 0.56) (Fig.~\ref{fig:cross-case}(a)). The same models therefore tracked a change in the PT without retraining in both cases, with a lower and wider range in the high tier. In the AIT-400, ARIMA (0.89) and the previous-run copy (0.92) matched or exceeded the best language model (0.89). The evidence supports tracking without retraining and does not support higher accuracy than classical in-context baselines.

Adaptation to the human was measured by the adaptation gain \(G_H\) on the persona-estimate error. Across the five drift scenarios (Fig.~\ref{fig:cross-case}(b)), the DDD-GenDT operator had a higher gain (1.30 to 1.51, median 1.45) than the AIT-400 trainee (1.16 to 1.34, median 1.23). The time-to-adapt was right-censored in 20\% to 77\% of the operator drift runs and 16\% to 72\% of the trainee drift runs, so the estimate often did not re-enter its tolerance band before the stream ended. PRISM estimates the learner state via zero-shot sentiment analysis (F1 of 0.91 for GPT-4) and has no error series, so \(G_H\) does not apply to it.

\subsection{Coupling}
Coupling was quantified as the ratio of the DT latency to the timescale of the process it serves (Fig.~\ref{fig:cross-case}(c)). For the PT step, the median ratio \(\rho\) was 55 in the medium tier (24.8 to 56.7) and 85 for the high-tier forecaster (39.5 to 86.7), so the DT latency exceeded the PT timescale by more than an order of magnitude for both forecasters. The vision model of the station required 57.2 ms (\(\rho\) of 0.035) and is the only component with a PT-step ratio below 1. The forecasters therefore cannot close the loop at the PT rate, and their outputs serve the human decision step. For the human step, the ratio \(\rho_H\) was 0.35 to 0.74 for the AIT-400 trainee, 1.14 to 1.18 for three of the four DDD-GenDT operator models (0.52 for Llama-3.1-8B-Instruct), and 0.20 to 1.34 for PRISM across five models and four step durations, of which only GPT-4 Turbo at the spoken learner duration exceeded 1 (1.34), and GPT-4 at that duration (0.95) and GPT-4 Turbo at the read tutor duration (1.18) were indeterminate. The human step is thus met in the high tier, met in the low tier in 17 of 20 model-duration combinations, and exceeded in the medium tier by most models. The tiers time different components and use different human timescales and adaptation metrics, so the values are compared within each panel rather than as a single score.

\section{Conclusion}
\label{sec:conclusion}

Digital twins of cyber-physical systems face three constraints: data are scarce, the physical asset changes after deployment, and different humans require different representations. This paper proposed HAMF-DT, a human-centric adaptive multi-fidelity DT framework that models the evolving human, the PT, and the DT core, and treats fidelity as a design dimension with low, medium, and high tiers. A human inference and adaptation function estimates the human state, predicts the response, and adapts the representation. The tiers are distinguished by two axes, namely the update of the representation in response to new data and the coupling ratio of the DT latency to the duration of the process it follows. Three instantiations, a virtual reality tutor (PRISM), a CNC milling model (DDD-GenDT), and an industrial training station (AIT-400), were evaluated along both axes.

The tiers differ by which functions adapt and where the latency constraint lies. The low-fidelity twin maintains a static PT representation and estimates learner sentiment with an accuracy of 0.86 (F1 of 0.91 for GPT-4), and its human-step ratio \(\rho_H\) ranges from 0.20 to 1.34. The medium-fidelity twin reduces its forecast error without retraining (\(D\) from 0.65 to 0.88) and lowers the persona-estimate error (\(G_H\) from 1.30 to 1.51), but its latency exceeds the PT timescale by 24.8 to 56.7 times (\(\rho\)) and the operator step for three of four models (\(\rho_H\) from 1.14 to 1.18). The forecaster of the high-fidelity station, a medium-tier component, shows smaller declines (\(D\) from 0.39 to 0.89) and gains (\(G_H\) from 1.16 to 1.34), a PT-step ratio \(\rho\) of 39.5 to 86.7, and a trainee-step ratio \(\rho_H\) of 0.35 to 0.74, whereas its vision model meets the PT step (\(\rho\) of 0.035). In the AIT-400, ARIMA and the previous-run copy matched or exceeded the best language model, so the evidence supports tracking without retraining and indicates that accuracy does not exceed classical in-context baselines. Gain and coupling therefore do not increase monotonically with the tier.

The three instantiations serve as proofs of concept. They use recorded PT data in the medium- and high-fidelity cases and simulate humans in place of operators and trainees. They time different components using different adaptation metrics. They show that the framework can be instantiated at each tier and that the tiers differ in adaptivity and coupling. They do not include a live round trip between the DT and the PT, so the coupling ratios describe the DT latency relative to the process step rather than the performance of a closed loop. Future work will evaluate the human layer with participants and measure coupling in a live closed loop for the high tier.

\section*{Acknowledgments}
This work was supported in part by the National Science Foundation (NSF) under Grant 2335046 and Grant OIA-2218046, and in part by the OpenAI Researcher Access Program under Grant 0000011862.

\section*{Use of Generative AI}
The authors used Claude Sonnet 5.0 and Claude Sonnet 5.5 (Anthropic) to improve grammar, rephrase text, and format tables and algorithms. The authors reviewed and edited all content and take full responsibility for it.

\bibliographystyle{unsrtnat}
\bibliography{references}

\end{document}